\documentclass[10pt,twocolumn,twoside]{IEEEtran}
\usepackage{amsmath,amsfonts}
\usepackage{algorithmic}
\usepackage{algorithm}
\usepackage{array}
\usepackage[caption=false,font=normalsize,labelfont=sf,textfont=sf]{subfig}
\usepackage{textcomp}
\usepackage{stfloats}
\usepackage{url}
\usepackage{verbatim}
\usepackage{graphicx}
\usepackage{cite}
\usepackage{aliascnt}
\usepackage{booktabs}
\usepackage{makecell}
\usepackage{array}
\usepackage{graphicx}
\usepackage{subcaption}
\usepackage{enumitem}
\usepackage[most]{tcolorbox}
\usepackage{kotex}  

\definecolor{apcback}{rgb}{0.862,0.902,0.973}
\definecolor{apcframe}{rgb}{0.353,0.490,0.729}

\usepackage{graphicx}
\begin{document}

\title{BERT-APC: A Reference-free Framework for Automatic Pitch Correction via Musical Context Inference}

\author{Sungjae Kim, Kihyun Na, Jinyoung Choi, and Injung Kim\footnote{Corresponding author.}
\thanks{The authors are with Computer Science and Electrical Engineering Department, Handong Global University, Pohang 37554, Korea. (e-mail: sjkim@handong.ac.kr; kevinna95@gmail.com; jinyoung@handong.ac.kr; ijkim@handong.edu)}
\thanks{Manuscript received November 18, 2025}}

\markboth{Journal of \LaTeX\ Class Files,~Vol.~14, No.~8, August~2021}%
{Kim \MakeLowercase{\textit{et al.}}: BERT-APC: A Reference-free Framework for Automatic Pitch Correction via Musical Context Inference}


\maketitle

\begin{abstract}
Automatic Pitch Correction (APC) enhances vocal recordings by correcting pitch deviations to align with the intended musical notes.
However, existing APC systems either rely on reference pitches, which limits their practical applicability, or employ simple pitch estimation algorithms that often fail to preserve expressiveness and naturalness. We propose BERT-APC, a novel reference-free APC framework that corrects pitch errors while maintaining the expressiveness and naturalness of vocal performances.
In BERT-APC, a novel stationary pitch predictor first estimates the stationary pitch of each note—defined as the continuous pitch derived from the stable region within a note and serving as an approximation for the perceived pitch—from the detuned singing voice.
Subsequently, a context-aware note pitch predictor infers the intended pitch sequence by leveraging a repurposed music language model that incorporates musical context.
Finally, a note-level correction algorithm fixes note-level pitch errors while preserving intentional pitch deviations for emotional expression.
In addition, we introduce a learnable data augmentation strategy that improves the robustness of the music language model by simulating realistic detuning patterns.
Compared to two recent singing voice transcription models, BERT-APC demonstrated superior performance in target note pitch prediction, outperforming the second-best model, ROSVOT, by 10.49 percentage points (pp) on highly detuned samples in terms of the raw pitch accuracy.
In the MOS test, BERT-APC achieved the highest quality rating of 4.32 $\pm$ 0.15, which is significantly higher than those of the widely-used commercial APC tools, Auto-Tune (3.22 $\pm$ 0.18) and Melodyne (3.08 $\pm$ 0.18), while maintaining a comparable ability to preserve expressive nuances.
To the best of our knowledge, this is the first APC model that leverages a music language model to achieve reference-free pitch correction with symbolic musical context. The corrected audio samples of BERT-APC are available online.\footnote{\url{https://joshua-1995.github.io/BERT-APC-Demo/}}
\end{abstract}


\section{Introduction}
\IEEEPARstart{A}\ utomatic Pitch Correction (APC) is a critical technique in modern music production that enhances vocal performance by correcting pitch errors. Recent deep learning-based APC systems~\cite{apc_deepautotuner20, apc_diffpitcher23, apc_contuner24} have demonstrated impressive pitch correction performance by leveraging external references, such as annotated music scores~\cite{apc_karatuner22, apc_diffpitcher23, apc_contuner24} or professionally tuned guide vocals ~\cite{apc_ctw,apc_nsvb22}. These references provide strong guidance, enabling precise corrections. However, the reliance on such references hinders their applications in many real-world scenarios, where such resources are often unavailable or costly to produce.

Widely used commercial APC systems, such as Auto-Tune~\cite{apc_autotune24} and Melodyne~\cite{apc_melodyne}, provide reference-free pitch correction based on rule-based or signal-processing techniques. Auto-Tune applies scale-constrained pitch quantization, adjusting detuned input pitches to the closest discrete pitches within a user-specified musical scale. 
Melodyne corrects pitches at the note level, enabling adjustments within musically coherent units and allowing better preservation of expressive variations such as vibrato and pitch glides. Beyond these commercial systems, there have also been signal-processing-based efforts to preserve expressive variations such as vibrato during pitch correction~\cite{apc_dpw}.
Although these systems operate without external references, they often neglect higher-level musical contexts—such as harmonic structure, tonal progression, and phrase-level coherence—which can lead to implausible pitch corrections that sound musically unnatural.

One possible strategy for reference-free APC is to leverage Singing Voice Transcription (SVT) models to extract discrete pitch sequences from input singing voices. Early SVT models estimate note pitches based on simple statistics such as median~\cite{svt_vocano21,svt_phonemesvt23}, which often replicate the pitch errors of the input audio onto the transcribed note pitches. 
Recent models~\cite{svt_transformer23, svt_rosvot24, svt_timealigned25, svt_stars25} predict discrete pitches using neural network classifiers, demonstrating improved robustness against moderate pitch deviations. However, they rely solely on acoustic features and do not exploit musical context, making them less reliable when pitch deviations are substantial.

To address these limitations, we propose BERT-APC, a novel reference-free APC model. BERT-APC leverages a music language model—originally developed for symbolic music understanding—to correct vocal pitch errors while maintaining consistency with the surrounding musical context.
Symbolic music language models~\cite{smu_musicbert21, smu_piano21, smu_pianobart24, smu_midibert24, smu_adversarial25}, trained on large collections of symbolic music data, have demonstrated strong capabilities in capturing patterns of harmony, tonality, and melodic flow. We repurposed a recent music language model, MusicBERT \cite{smu_musicbert21}, to provide context-aware guidance for APC, supplementing the acoustic features of the input audio.
Even without ground-truth (GT) references, our method enables the estimation of plausible and musically coherent target pitches, effectively resolving ambiguities in cases of highly detuned singing voices. 

Incorporating symbolic language models into an APC system poses challenges due to the modality mismatch between continuous vocal pitches and the discrete input representation of the symbolic language models. To correct pitch error using a symbolic language model, the input audio must be segmented into notes, and the pitch of each note segment must be quantized. However, the presence of transitional regions between notes and vocal ornamentations-such as vibrato and pitch glides-often blurs note boundaries and introduces substantial variations in pitch, thereby hindering accurate note segmentation and note-level pitch estimation. To address these challenges, we present a deep learning-based Note Segmentator (NS) along with a Stationary Pitch Predictor (SPP), which together estimate the perceived pitch of each note despite the presence of ambiguous pitch patterns.

An additional advantage of the proposed BERT-APC is its ability to preserve subtle pitch variations that are intentionally introduced for expressive purposes. Because BERT-APC performs pitch correction at the note level, it is able to retain fine-grained variations at the frame level, thereby maintaining musical continuity and naturalness. Compared with two recent SVT models—PhonemeSVT~\cite{svt_phonemesvt23} and ROSVOT~\cite{svt_rosvot24}—BERT-APC achieved substantially higher target note pitch prediction accuracy, outperforming them by 33.59 and 10.49 percentage points (pp), respectively, in terms of Raw Pitch Accuracy (RPA) metric~\cite{metric_rpa14} on the highly detuned test subset. Furthermore, in a 5-point quality MOS evaluation of pitch-corrected audio, BERT-APC achieved significantly higher pitch accuracy (4.32 ± 0.15) than two commercial APC systems, Auto-Tune (3.22 ± 0.18) and Melodyne (3.08 ± 0.18), while maintaining a comparable ability to preserve expressive nuances.


To the best of our knowledge, BERT-APC is the first reference-free APC model that leverages a symbolic music language model for pitch correction.
The contributions of our work are summarized as follows:
\begin{itemize}
    \item A novel reference-free APC framework, BERT-APC, which leverages a musical language model to correct detuned vocal pitches by incorporating musical context.
    \item A neural note segmentator that segments singing voices with diverse variations into discrete notes.
    \item A stationary pitch predictor designed to estimate the perceived pitch of each note even when the input pitch sequence includes transitions and vocal ornamentations.
    \item A learnable detuner for data augmentation, designed to enhance APC models by injecting pitch deviations derived from real-world detuning patterns in off-pitch singing voices.
\end{itemize}







\section{Related Work}
\subsection{Automatic Pitch Correction}
Previous studies on APC can be broadly divided into reference-based APC, which utilizes reference pitches from music scores, instrumental accompaniments, and guide vocals, and reference-free APC, which corrects out-of-tune singing voices without relying on any reference.

\subsubsection{Reference-based APC}
Deep AutoTuner~\cite{apc_deepautotuner20} estimates pitch shifts from the joint spectral input of the vocal and its time-aligned accompaniment, implicitly promoting harmonic compatibility between them. KaraTuner~\cite{apc_karatuner22} employs a Transformer-based pitch predictor conditioned on aligned note pitches to generate the pitch contour and synthesizes pitch-corrected voices through a pitch-controllable neural vocoder. Diff-Pitcher~\cite{apc_diffpitcher23} employs a vocal-adaptive pitch predictor to estimate the output pitch contour and further refines it using a diffusion-based model, producing high-quality and natural-sounding voice signals matched to the target pitch. 
More recently, ConTuner~\cite{apc_contuner24} predicts expressive pitch contours from note-level inputs and spectral features via an expressiveness enhancer trained on amateur-professional vocal pairs.
These reference-based models generally achieve accurate and natural-sounding corrections, but their reliance on reference materials limits their applicability in real-world settings. In contrast, our work focuses on pitch correction in a fully reference-free setting.

\subsubsection{Reference-free APC}
Commercial tools such as Auto-Tune \cite{apc_autotune24} and Melodyne \cite{apc_melodyne} are widely adopted due to their simplicity and capability for real-time control. Operating without external references, these systems map the input pitch to the nearest discrete pitch within a predefined scale (e.g., 12-tone equal temperament). However, such a simplistic quantization approach fails to consider broader musical contexts, including harmonic progressions and phrase structures, which may result in musically unnatural corrections.

\subsection{Singing Voice Transcription}
Singing Voice Transcription (SVT) is the task of automatically transforming vocal performances into symbolic musical representations, including pitch contours, note onset times, durations, and corresponding lyrics.
Most SVT models focus on note boundary detection and derive note pitches based on simple statistical features, such as the median~\cite{svt_vocano21} or the weighted median~\cite{svt_phonemesvt23}.
As they do not explicitly infer the intended note pitch, these models lack the ability to recover pitch information from off-key or out-of-tune singing voices.

Recent SVT models~\cite{svt_transformer23, svt_rosvot24, svt_timealigned25} have introduced pitch classification networks that directly predict discrete pitches from the input audio.
However, these models still rely solely on acoustic features and lack awareness of musical context, making them unreliable for highly deviated singing voices or inputs with ambiguous pitches.

\subsection{Language Models for Symbolic Music Understanding}
Symbolic Music Understanding (SMU) refers to the process of analyzing and interpreting music represented in a symbolic form—such as MIDI files and music scores—rather than as raw audio signals.
Recently, symbolic music language models such as MusicBERT~\cite{smu_musicbert21}, MidiBERT~\cite{smu_midibert24}, and Adversarial-MidiBERT~\cite{smu_adversarial25}, trained on large collections of discrete musical tokens have demonstrated their effectiveness in capturing high-level musical contexts, including harmonic relationships, tonality, and melodic structures.

While these music language models are primarily used for music analysis, retrieval, and generation tasks, in this study, we repurposed one of them to estimate the intended discrete pitch from detuned vocal pitches.

\section{BERT-APC}
\begin{figure}[t]
  \centering
  \hspace*{-0.03\textwidth}
  \includegraphics[width=0.5\textwidth]{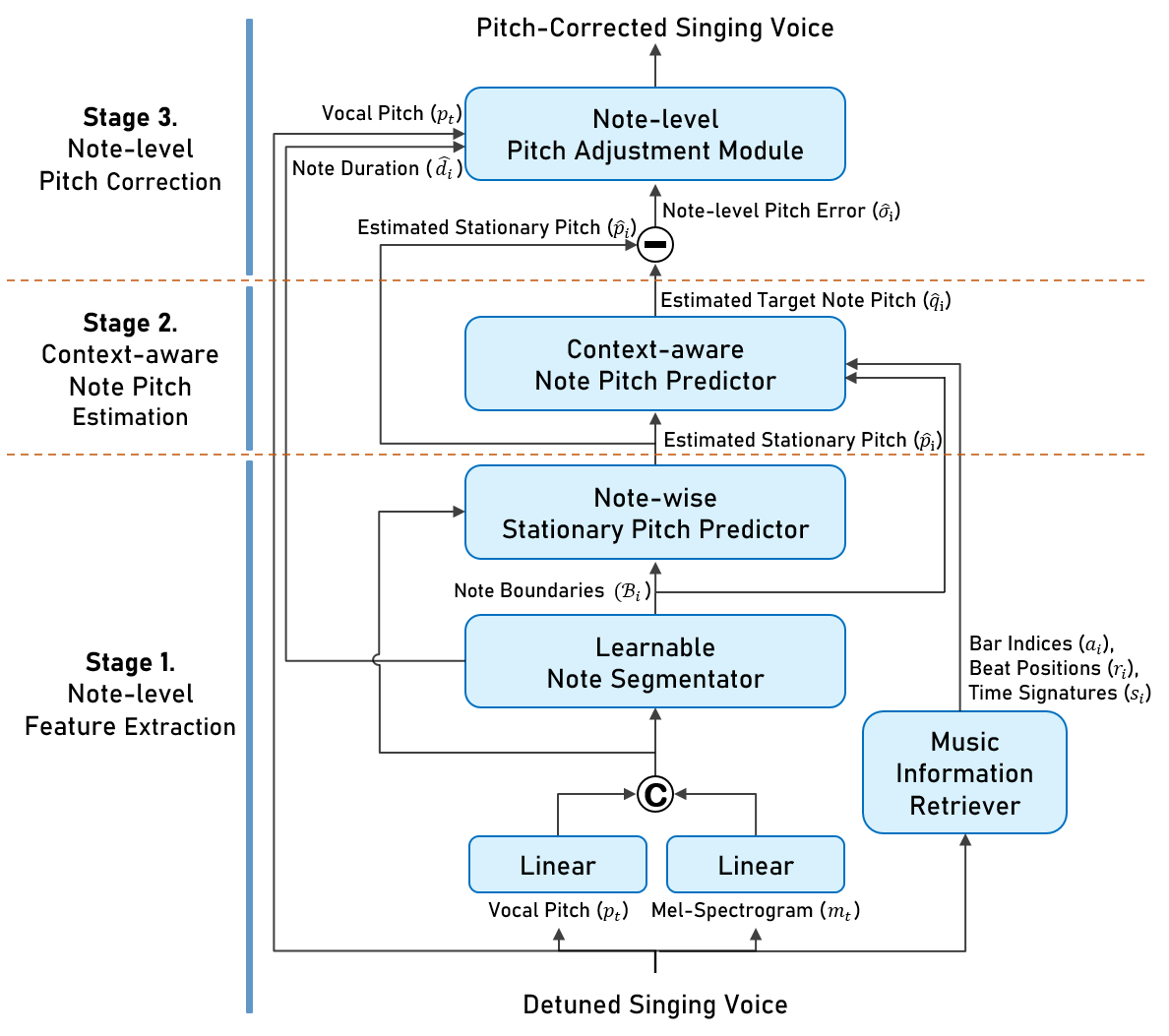}
  \caption{\textbf{Model architecture of BERT-APC}. The system operates in three stages—note-level feature extraction, context-aware note pitch estimation, and note-level pitch correction. A concise step-by-step overview is provided in Fig.~\ref{fig:step_by_step_overview}.}
  \label{fig:model_architecture}
\end{figure}

\subsection{Overview}

BERT-APC corrects detuned singing voices through a three-stage process, as illustrated in Fig.~\ref{fig:model_architecture}. First, BERT-APC extracts note-level features from the input singing voice via a note segmentator and a stationary pitch predictor. Then, it predicts the intended note pitches by leveraging a repurposed music language model, MusicBERT. Finally, BERT-APC corrects the pitch deviation at the note level while preserving the expressive characteristics of the singing voice that convey emotional nuances. The main technical contribution of this work lies in the first two stages and in the construction of the frame-level target pitch contour in the third stage. To modify the pitch of the audio signal to match the estimated target pitch contour, we adopt a time-domain pitch-synchronous overlap-add (TD-PSOLA) algorithm implemented in the Praat-Parselmouth library~\cite{parselmouth}.


For clarity, we define the pitch-related terms and the corresponding notations used in this paper. $t$ and $i$ denote the frame and note indices, respectively. The \emph{input vocal pitch}, denoted by $p_t$, is the frame-level continuous pitch extracted from the input singing voice. The \emph{stationary pitch}, $\hat{p}_i$, is defined as the continuous pitch estimated from the stable region within a note and serves as an approximation of the perceived pitch.
The \emph{GT note pitch}, $q_i$, refers to the discrete pitch specified in the musical score that the singer intends to produce, and its estimate, $\hat{q}_i$, is referred to as the \emph{estimated target note pitch}. The \emph{target pitch} refers to the pitch that pitch adjustment aims to achieve. Ideally, the target note pitch corresponds to the GT note pitch $q_i$, and the frame-level target pitch, $p^*_t$, consists of the GT note pitch with added expressive frame-level variations; however, in practice, their estimates are used during the pitch correction process. Throughout this paper, all pitch values are expressed on a semitone scale.




\subsection{Note-level Feature Extraction}

BERT-APC leverages a symbolic music language model to estimate the intended note pitches and to correct pitch deviations at the note level.
To achieve this, BERT-APC first segments the frame-level features of the input singing voice into note-level units, and then estimates a stationary pitch $\hat{p}_i$ for each note.

Both the note segmentator and the stationary pitch predictor consist of a combination of a Transformer encoder and a prediction head, respectively. Their encoders, $\mathcal{E}_{ns}(\cdot)$ and $\mathcal{E}_{spp}(\cdot)$, share the same architecture but differ in parameters.
The encoder inputs are formed by concatenating the vocal pitch sequence $p \in \mathbb{R}^{T}$ and a Mel-spectrogram $m \in \mathbb{R}^{T \times C}$, where $T$ and $C$ denote the numbers of frames and channels, respectively. The concatenated inputs are encoded into a hidden representation $h \in \mathbb{R}^{T \times D}$, as Eq. (\ref{eq:encoder}), where $* \in \{ns, spp\}$:
\begin{equation}
    h_* = \mathcal{E}_*(Concat(p,m)).
    \label{eq:encoder}
\end{equation}

\begin{figure}[t]
\centering
\begin{tcolorbox}[
    colback=apcback,
    colframe=apcframe,
    title=Step-by-Step Overview,
    fonttitle=\bfseries,
    left=1mm, right=1mm, top=1mm, bottom=1mm,
    boxrule=0.4pt
]
\footnotesize
\textbf{Stage 1: Note-level Feature Extraction} \\ 
$\triangleright$ Extract frame-level acoustic features, vocal pitch and Mel-spectrogram.\\
$\triangleright$ Encode the input features into hidden representation (Eq.~\ref{eq:encoder}).\\
$\triangleright$ Estimate note boundaries via the \emph{note segmentator} (Eq.~\ref{note_segmentator}, Alg.~\ref{alg:nms}).\\
$\triangleright$ Compute stationary pitch for each note interval via the \emph{note-wise stationary pitch predictor} (Eq.~\ref{eq:stationary_pitch_estimator}, \ref{eq:learnable_weight_estimator}).\\[4pt]

\textbf{Stage 2: Context-aware Note Pitch Estimation}\\
$\triangleright$ Predict note pitches for the pitch correction target via \emph{context-aware note pitch predictor} (Eq.~\ref{eq:cnpp_out}). \\[4pt]

\textbf{Stage 3: Note-level Pitch Correction}\\
$\triangleright$ Compute the note-level pitch error by subtracting the estimated target note pitch from the estimated stationary pitch (Eq.~\ref{eq:pitch_error}).\\
$\triangleright$ Compute the target pitch for each frame within a note by subtracting the corresponding note-level pitch error from the input vocal pitch. \\
$\triangleright$ Adjust the pitch of the detuned singing voice to match the target pitch via the \emph{pitch adjustment module}.
\end{tcolorbox}
\caption{Overview of the three-stage process used by BERT-APC for reference-free pitch correction.}
\label{fig:step_by_step_overview}
\end{figure}

\subsubsection{Note Segmentator}
The head of the note segmentator $NS(\cdot)$ estimates boundary probability for each frame from the encoder output, as in Eq.~(\ref{note_segmentator}), where $\hat{b} = (b_1, \ldots, b_T)$, and $b_t$ denotes the probability that the $t$-th frame is a note boundary.
\begin{equation}
    \hat{b} = NS(h_{ns})
    \label{note_segmentator}
\end{equation}
In this study, we implemented the note segmentator head by combining a GRU and a linear layer with the Sigmoid activation.

Since non-boundary frames overwhelmingly outnumber boundary frames, we employed focal loss~\cite{focal_loss} to improve robustness against this class imbalance.
The boundary label of a training example consists of a binary vector $y_b=(y_1,...,y_T), y_t \in \{0, 1\}$, where $y_t = 1$ indicates that the $t$-th frame is a boundary. As identifying the exact temporal positions of note boundaries from frame-level features is challenging, we convert the hard label into the corresponding soft label $\tilde{y}_b = (\tilde{y}_1,...,\tilde{y}_T), \tilde{y}_t \in [0, 1]$ by a Gaussian kernel, following common practice in SVT models~\cite{svt_rosvot24}. The training objective of the note segmentator is presented in Eq. (\ref{eq:note_segmentator_loss}). Here, $\gamma$ denotes the focusing parameter, and $\alpha_t$ is a weighting factor used to address the imbalance between boundary and non-boundary frames. We set $\gamma = 4$, and $\alpha_t = 1$ for non-boundary frames and 29 for boundary frames.

\begin{align}
\mathcal{L}_{\text{boundary}}
&= -\sum_{t=1}^{T} \alpha_t \big[
(1 - \hat{b}_t)^{\gamma} \tilde{y}_b(t) \log \hat{b}_t \nonumber \\
&\quad + \hat{b}_t^{\gamma} (1 - \tilde{y}_b(t)) \log (1 - \hat{b}_t)
\big].
\label{eq:note_segmentator_loss}
\end{align}

\begin{algorithm}[t]
\small
\caption{Greedy NMS for Boundary Detection}
\label{alg:nms}
\begin{algorithmic}[1]
  \STATE \textbf{Input:} frame-wise boundary prob. $\hat{b}$, window $w$, threshold $\theta$
  \STATE $\mathcal{B} \leftarrow \emptyset$, $\tilde{b} \leftarrow \hat{b}$
  \WHILE{$\max \tilde{b} \ge \theta$}
    \STATE $t^* \leftarrow \arg\max_t \tilde{b}_t$
    \STATE $\mathcal{B} \leftarrow \mathcal{B} \cup \{t^*\}$
    \STATE $\tilde{b}[t^*-w : t^*+w] \leftarrow 0$.
  \ENDWHILE
  \STATE Add the first and last frames of the singing region to $\mathcal{B}$
  \STATE Sort $\mathcal{B}$ in ascending order
  \RETURN $\mathcal{B}$
\end{algorithmic}
\end{algorithm}

To detect boundary frames from the frame-level boundary probabilities $\hat{b}$, we apply a greedy non-maximum suppression (NMS) with minimum distance $w$ and threshold $\theta$, as Alg.~\ref{alg:nms}.

In our implementation, we set $w = 5$, corresponding to a $\pm 58$ msec window given a sampling rate of 22,050 Hz and a hop size of 256 samples. This NMS algorithm retains only the most salient and temporally distinct boundaries, improving segmentation quality and avoiding spurious boundary clutter.

\begin{figure*}[t]
\centering
\begin{minipage}{0.30\linewidth}
  \centering
  \includegraphics[width=\linewidth]{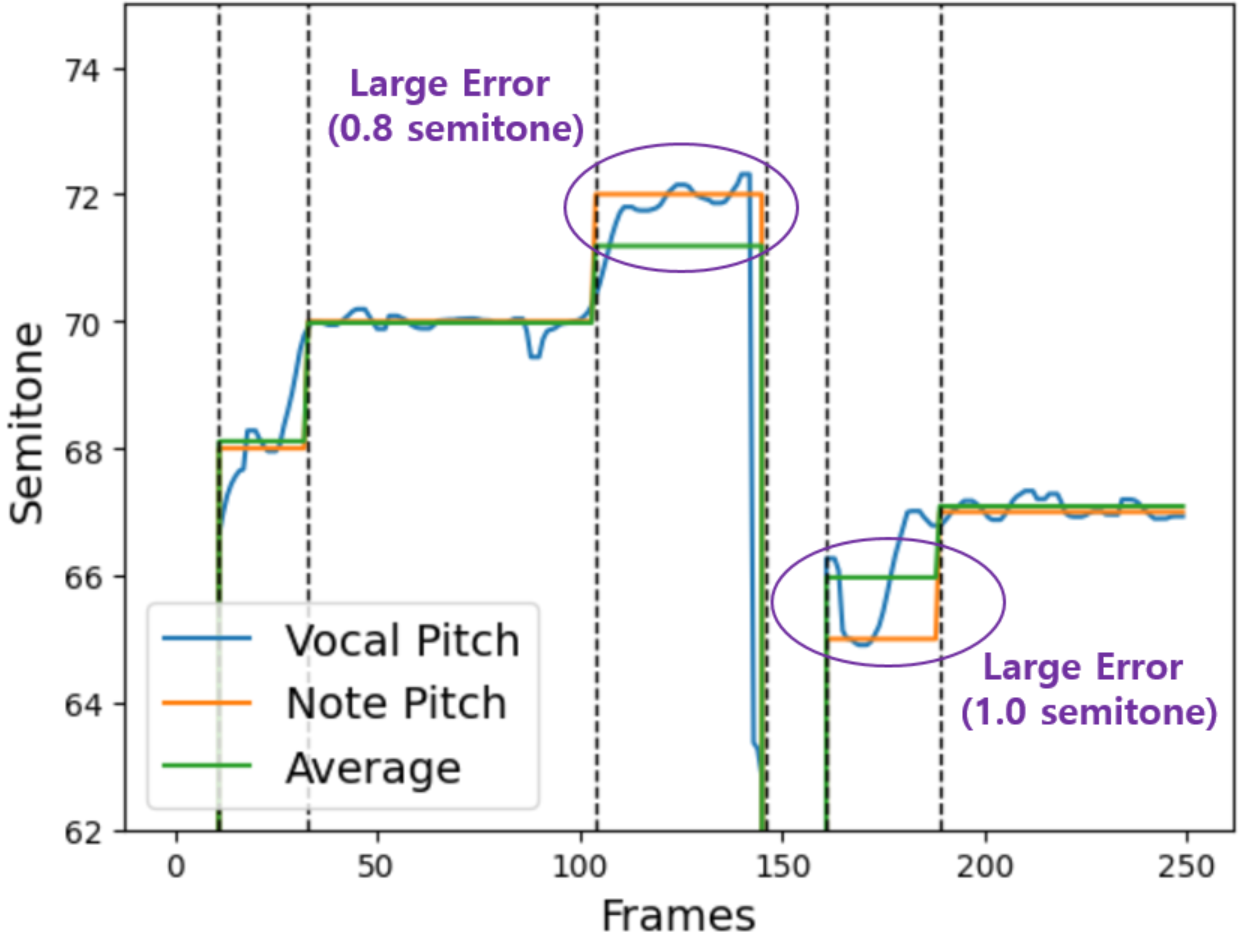}
  \caption*{(a) Average~\cite{moore2012introduction}}
\end{minipage}\hfill
\begin{minipage}{0.30\linewidth}
  \centering
  \includegraphics[width=\linewidth]{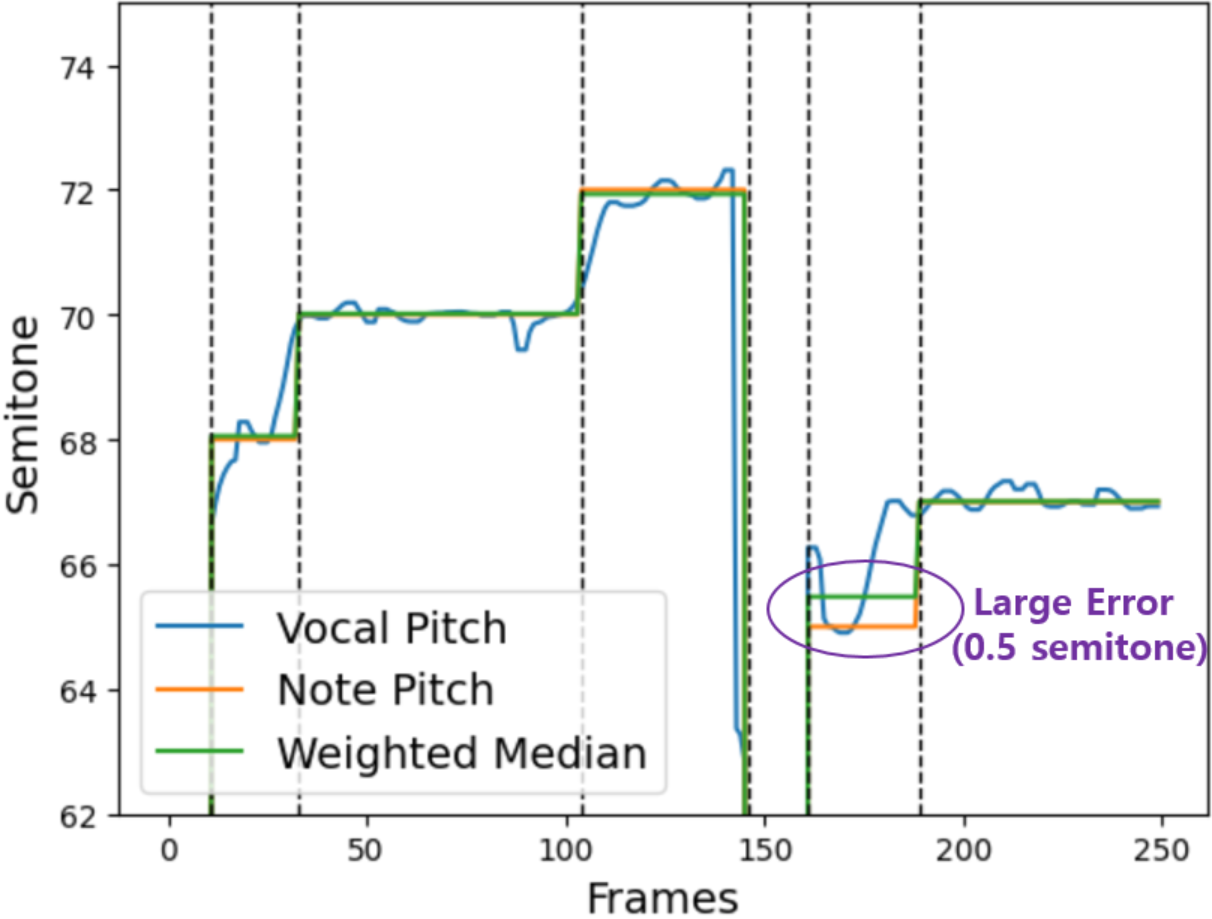}
  \caption*{(b) Weighted Median~\cite{svt_phonemesvt23}}
\end{minipage}\hfill
\begin{minipage}{0.30\linewidth}
  \centering
  \includegraphics[width=\linewidth]{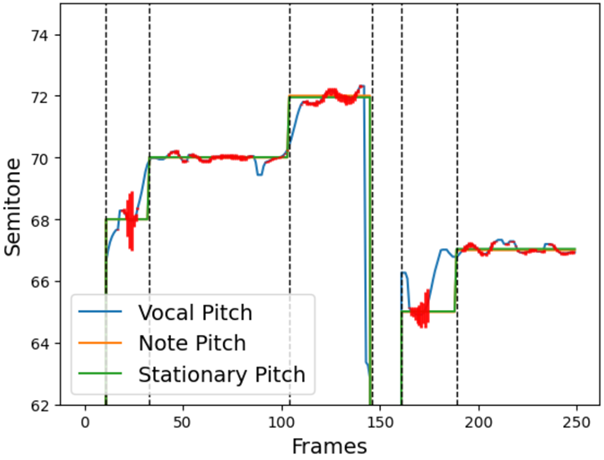}
  \caption*{(c) Stationary Pitch Predictor (Ours)}
\end{minipage}

\caption{\textbf{Comparison of pitch estimation methods.} Blue, orange, and green lines denote vocal, ground-truth, and estimated pitches, respectively. The estimation methods based on (a) average and (b) weighted median exhibit substantial error for the notes with large pitch variations, as marked by purple ellipses. In contrast, the proposed method successfully identifies perceptual pitch centers, demonstrating robustness against pitch variation, as shown in (c). The height of the red regions visualizes estimated stationarity weight $w_t$ (Eq.~\ref{eq:learnable_weight_estimator}) for each frame. 
Notably, in the vibrato segment on the right, frames near the mid-pitch region receive relatively high weights, suggesting that the proposed method can estimate the perceived pitch center even for notes with highly varying pitches.
}
\label{fig:pitch_comparison}
\end{figure*}

\subsubsection{Note-wise Stationary Pitch Predictor}
To correct pitch errors in a singing voice, it is necessary to identify the pitch of each note. However, the pitch contour of a singing voice contains not only the note pitches but also various fluctuations, such as inter-note transitions, vocal ornamentations used for expressive purposes, and pitch errors. As a result, determining a representative pitch for each note is challenging.

Previous work has shown that the pitch perceived by listeners from a singing voice primarily corresponds to the pitch of the stationary regions, while segments with fluctuations such as vibrato are perceived in terms of their average pitch~\cite{moore2012introduction}.
Yong et al. proposed a weighted median approach that assigns higher weights to frames near the center of the note using the Hann window \cite{svt_phonemesvt23}. This method performs well when a clear stationary region exists in the center, but fails when transitions are asymmetric or the stationary region is off-center, as shown in Fig.~\ref{fig:pitch_comparison}(b). 

A commercial tool, Melodyne 5, estimates the pitch center of a note using a musically weighted algorithm that assigns higher weights to perceptually salient and stable regions, while down-weighting fluctuating segments such as vibrato or drift~\cite{apc_melodyne}. However, its detailed procedure has not been disclosed to the public.

To reliably estimate $\hat{p}_i$ in the presence of diverse variations, we developed a learnable stationary pitch predictor. It estimates $\hat{p}_i$ as a weighted average of frame-level pitches within the note interval, as Eq. (\ref{eq:stationary_pitch_estimator}):
\begin{equation}
    \hat{p}_i=\sum_{t\in I(i)}{w_t p_t},
    \label{eq:stationary_pitch_estimator}
\end{equation}
where $I(i)$ denotes the interval of the $i$-th note and $w_t$ is the corresponding weight for frame $t$. While previous studies determined weights via handcrafted algorithms,
we propose a learnable weight predictor $\operatorname{WP}(\cdot)$ as Eq. (\ref{eq:learnable_weight_estimator}).
\begin{align}
    \label{eq:learnable_weight_estimator}
     e &= \operatorname{WP}(h_{spp})
\\
    \{ w_t \}_{t \in I(i)} &= \operatorname{Softmax}(\{e_t\}_{t \in I(i)})  \nonumber
\end{align}    
where $h_{spp} \in \mathbb{R}^{T \times D}$ and $e \in \mathbb{R}^{T}$.

In this study, we implemented the weight predictor by combining the Transformer encoder $\mathcal{E}_{spp}(\cdot)$ with a prediction head consisting of a single linear layer. 

A critical challenge in training the weight predictor lies in the difficulty of obtaining ground truth for stationary regions or frame-level weights $w_t$ at a low cost. To address this challenge, we train the weight predictor such that the estimated stationary pitch of each note, computed from in-tune training samples using Eq. (\ref{eq:stationary_pitch_estimator}), matches the GT note pitch $q_i$, i.e., $\hat{p}_i = q_i$ for each note $i$ in in-tune training samples. Since this condition is ill-posed, we add three regularization terms.
Consequently, the training objective of the stationary pitch predictor is presented in Eq. (\ref{eq:stationary_pitch_predictor_loss}):
\begin{align}
    \label{eq:stationary_pitch_predictor_loss}
    \mathcal{L}_{\text{spp}} &= \mathcal{L}_{\text{pitch}} + \lambda_s \mathcal{L}_{\text{stat}} + \lambda_d\mathcal{L}_{\text{dist}} + \lambda_u \mathcal{L}_{\text{uni}} \\
    &\mathcal{L}_{\text{pitch}} = \sum_i (\hat{p}_i - q_i)^2, \text{       }
    \mathcal{L}_{\text{stat}} = \sum_{t=1}^T \sigma_t^2 \cdot w_t \nonumber \\
    &\mathcal{L}_{\text{dist}} = \sum_{t=1}^T (p_t - q_i)^2 w_t,
    \mathcal{L}_{\text{uni}} = \sum_{t=1}^T w_t \log w_t. \nonumber
\end{align}

The primary regression loss, $\mathcal{L}_{\text{pitch}}$, measures the discrepancy between the estimated stationary pitch $\hat{p}_i$ and the GT note pitch $q_i$. The term $\mathcal{L}_{\text{stat}}$ penalizes high weights assigned to frames with large local pitch variations, where $\sigma_t^2$ denotes the local pitch variance around frame $t$, computed with a window size proportional to the note duration (10\% in our implementation). Smaller $\sigma_t^2$ values indicate more stable pitch regions, whereas larger values indicate fluctuating regions such as transitions or vibrato. $\mathcal{L}_{\text{dist}}$ penalizes high weights on frames with substantial pitch deviations. Additionally, $\mathcal{L}_{\text{uni}}$ promotes a smoother distribution of weights $w_t$ by increasing entropy. In our experiments, we set the hyperparameters as $\lambda_s = 0.05$, $\lambda_d = 0.1$, and $\lambda_u = 0.01$.

Fig.~\ref{fig:pitch_comparison} compares the results of the proposed stationary pitch predictor with two alternative pitch estimation methods, thereby demonstrating the effectiveness of our approach. In particular, despite its name, the \textit{stationary pitch predictor}, Fig.~\ref{fig:pitch_comparison}(c) suggests that the proposed method has the potential to successfully estimate the perceptual pitch center even in segments with highly-varying pitches.

\begin{figure}[t]
    \centering
    \includegraphics[width=\linewidth]{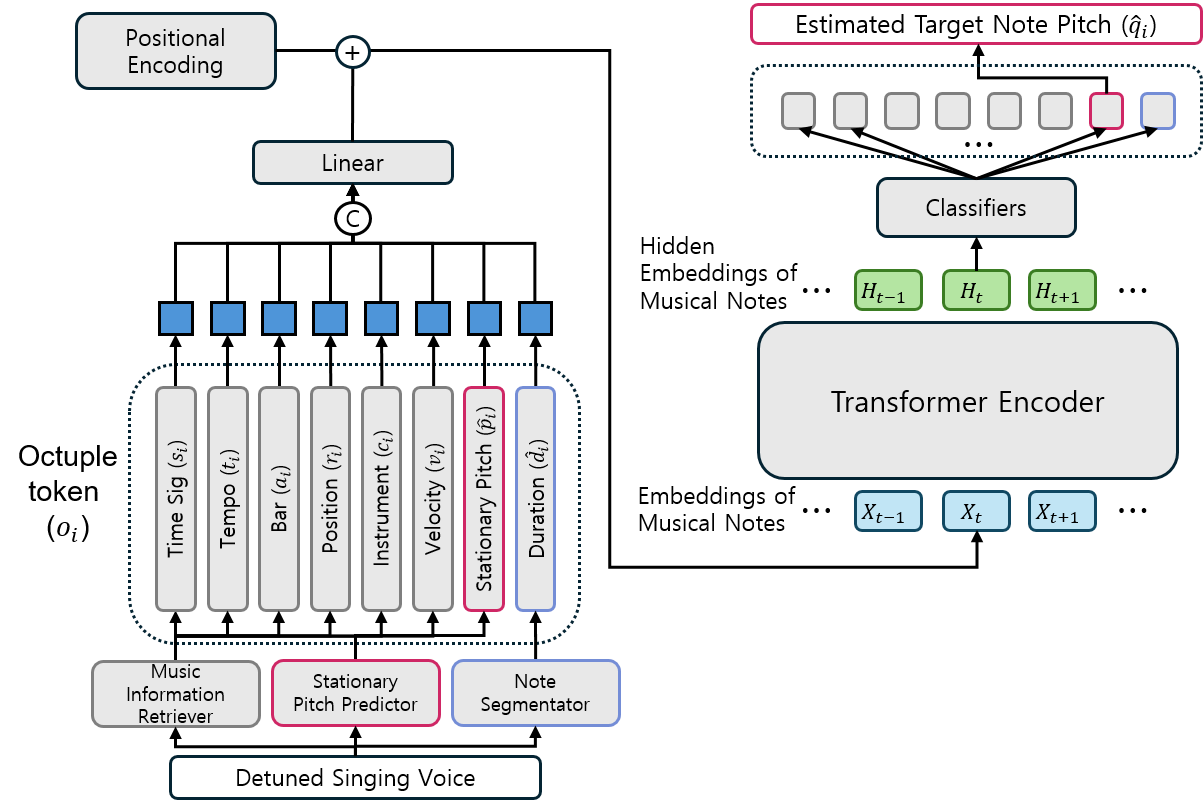}
    \caption{The architecture of the context-aware note pitch predictor that is based on the symbolic music language model, MusicBERT.}
    \label{fig:CNPP_architecture}
\end{figure}
\subsection{Context-aware Note Pitch Estimation}

The core challenge in reference-free APC is the accurate prediction of the target note pitch without relying on explicit melodic references. While the input singing voice alone may suffice when pitch errors are small (e.g., less than a semitone), severe pitch deviations present significant challenges. In such cases, acoustic features alone are insufficient to estimate the intended note pitch, necessitating additional context from high-level musical priors—such as tonal structure, melodic contour, and underlying chordal relationships among neighboring notes. Therefore, in Stage 2, BERT-APC estimates the intended note pitch by incorporating high-level musical priors beyond acoustic features alone.

To address this, we propose a \textit{Context-aware Note Pitch Predictor (CNPP)} that incorporates musical context into note-level pitch prediction. CNPP takes a sequence of detuned note segments and the corresponding fractional stationary pitch estimates as inputs, and predicts musically coherent target note pitches. It leverages a symbolic language model, MusicBERT~\cite{smu_musicbert21}, originally developed for music understanding, to infer musically plausible note pitches.

A technical issue in repurposing a symbolic language model for note pitch prediction is bridging the modality gap: the stationary pitches of the input voice have continuous values, whereas MusicBERT expects discrete symbolic tokens as input. A straightforward solution would be to round each pitch value to the closest discrete pitch; however, this naive approach leads to precision loss, preventing the symbolic model from capturing essential acoustic nuances and thus limiting its prediction performance. Therefore, we represent stationary pitches as \textit{interpolated pitch embeddings}. MusicBERT represents discrete pitches using learned embeddings, and we leverage these embeddings to represent stationary pitches. Specifically, we encode a stationary pitch by interpolating between the embeddings of the two closest discrete pitches, as in Eq.~\ref{eq:pitch_embedding_interpolation}:

\begin{equation}
\operatorname{interp}(\hat{p})
=
(1-\alpha)\cdot\operatorname{embed}\!\left(\lfloor \hat{p} \rfloor\right)
+ \alpha\cdot\operatorname{embed}\!\left(\lfloor \hat{p} \rfloor + 1\right)
\label{eq:pitch_embedding_interpolation}
\end{equation}
where $\alpha = \hat{p} - \lfloor \hat{p} \rfloor$ is the fractional part of the stationary pitch, $\operatorname{interp}(\cdot)$ denotes the pitch-embedding interpolation function, and $\operatorname{embed}(\cdot)$ is the learned embedding table. This method effectively represents fractional pitch values while maintaining representational compatibility with discrete pitches.

CNPP takes symbolic inputs represented as a sequence of octuple encodings $o_i$, as illustrated in Fig.~\ref{fig:CNPP_architecture}:
\begin{equation}
    o_i = (s_i, t_i, a_i, r_i, c_i, v_i, \hat{p}_i, \hat{d}_i)
    \label{eq:octuple}
\end{equation}
where $s_i$ is the time signature, $t_i$ is the tempo, $a_i$ is the bar index, $r_i$ is the beat position within the bar, $c_i$ is the instrument ID, $v_i$ is the velocity, $\hat{p}_i$ is the stationary pitch of the note, and $\hat{d}_i$ is the note duration.
While MusicBERT requires an instrument ID defined by the MIDI standard, BERT-APC is a reference-free APC model that uses only vocal-derived note events. Therefore, we assign a constant instrument ID (`Grand Piano') to all notes.
Among these, $\hat{p}_i$ and $\hat{d}_i$ are obtained from the stationary pitch predictor and the note segmentator, while $s_i, t_i, a_i,$ and $r_i$ are extracted from the input audio using the music information retrieval library Madmom~\cite{madmom}. CNPP outputs refined octuple encodings $\tilde{o}_i$ with the same format as $o_i$, from which we retrieve the estimated target note pitch $\hat{q}_i$:
\begin{align}
    \label{eq:cnpp_out}
    \hat{q}_i &= \operatorname{CNPP}(o_i)
\end{align}

We train the model using a cross-entropy loss between the predicted pitch token $\hat{q}_i$ and the corresponding ground-truth note pitch $q_i$.


We adopt MusicBERT, pretrained on a large corpus of symbolic music data, as the backbone of CNPP and fine-tune it to predict ground-truth note pitches from detuned pitch sequences. Fine-tuning CNPP requires a large number of detuned vocal pitch sequences and their corresponding GT note pitches. However, the quantity of highly detuned samples in public datasets is limited. Therefore, we applied a data augmentation technique to generate additional detuned samples.

While previous studies have synthesized detuned pitch sequences by adding random shifts to vocal pitches~\cite{apc_deepautotuner20, apc_enhancing23}, such a simple approach fails to capture the complex detuning patterns in real singing voices. Therefore, we developed a learnable detuner to simulate realistic detuning patterns. The proposed detuner takes a sequence of note pitches and durations as input and autoregressively predicts note-level pitch errors, which are added to the input pitches of CNPP to generate detuned pitch sequences. In this study, we implemented the detuner with a GRU module.

To train the detuner, we set the target note-level pitch error, $\epsilon_i$, to the pitch error measured from highly detuned samples, i.e., those within the top 10\% in terms of average pitch error. Specifically, $\epsilon_i$ is computed as the difference between the mean of the 30--70th percentile of $\{p_t\}_{t \in I(i)}$ and the GT note pitch $q_i$. 
The detuner is then trained to approximate $\epsilon_i$ by minimizing an MSE loss. Consequently, the proposed detuner learns to reproduce the detuning patterns observed in real low-quality singing voices.
At each training iteration, we stochastically detune each training sample with probability $p_{det}$. To maintain training stability, we adopt an annealing strategy that initializes $p_{det}$ at 0 and gradually increases it to 0.4.

\subsection{Note-level Pitch Correction via Pitch Adjustment Module} 


In Stage 3, the note-level pitch adjustment module adjusts the pitch of an input detuned singing voice to produce a pitch-corrected singing voice. This stage consists of three steps: 1) note-level pitch error computation, 2) frame-level target pitch computation, and 3) pitch manipulation.

\subsubsection{Note-level Pitch Error Computation}

Given the estimated stationary pitch $\hat{p}_i$ from Stage 1 and the predicted target note pitch $\hat{q}_i$ from Stage 2 for each note, we compute the note-level pitch errors as
\begin{equation}
\hat{\delta} = \{\hat{\delta}_i \mid \hat{\delta}_i = \hat{p}_i - \hat{q}_i\}_{i=1}^{N},
\label{eq:pitch_error}
\end{equation}
where $N$ is the number of notes.

\subsubsection{Frame-level Target Pitch Computation}

Using the note-level pitch error $\hat{\delta}_i$ and the corresponding note interval $I(i)$, the frame-level target pitch $p_t^*$ is computed as
\begin{equation}
p_t^* = p_t - \hat{\delta}_i,\quad t \in I(i),
\label{eq:target_pitch_contour}
\end{equation}
Because the same pitch shift is applied to all frames within each note, the relative intra-note pitch variations, such as vibrato, pitch bends, and portamento, are preserved.

\subsubsection{Pitch Adjustment}
Finally, we adjust the pitch contour of the input singing voice to match the frame-level target pitch, $p_t^*$.
With the estimated frame-level target pitch, the pitch of a singing voice can be adjusted using any pitch adjustment module. In this study, we adopt a TD-PSOLA-based pitch manipulation method implemented in the Praat-Parselmouth speech processing library, as it provides accurate pitch control while being simple, fast, and training-free. We compare Praat-Parselmouth with a modern pitch-controllable neural vocoder, SiFiGAN in Table~\ref{tab:pitch_adjustment_obj}.

Our correction algorithm maintains subtle fluctuations for expressive purposes by uniformly shifting the frame pitches within each note segment. In theory, large discrepancies in the degree of correction between adjacent notes may introduce discontinuities. However, such cases were not observed in our experiments.




\section{Experiments}
\begin{figure*}[t]
\centering

\begin{minipage}{0.32\textwidth}
    \centering
    \includegraphics[width=\linewidth]{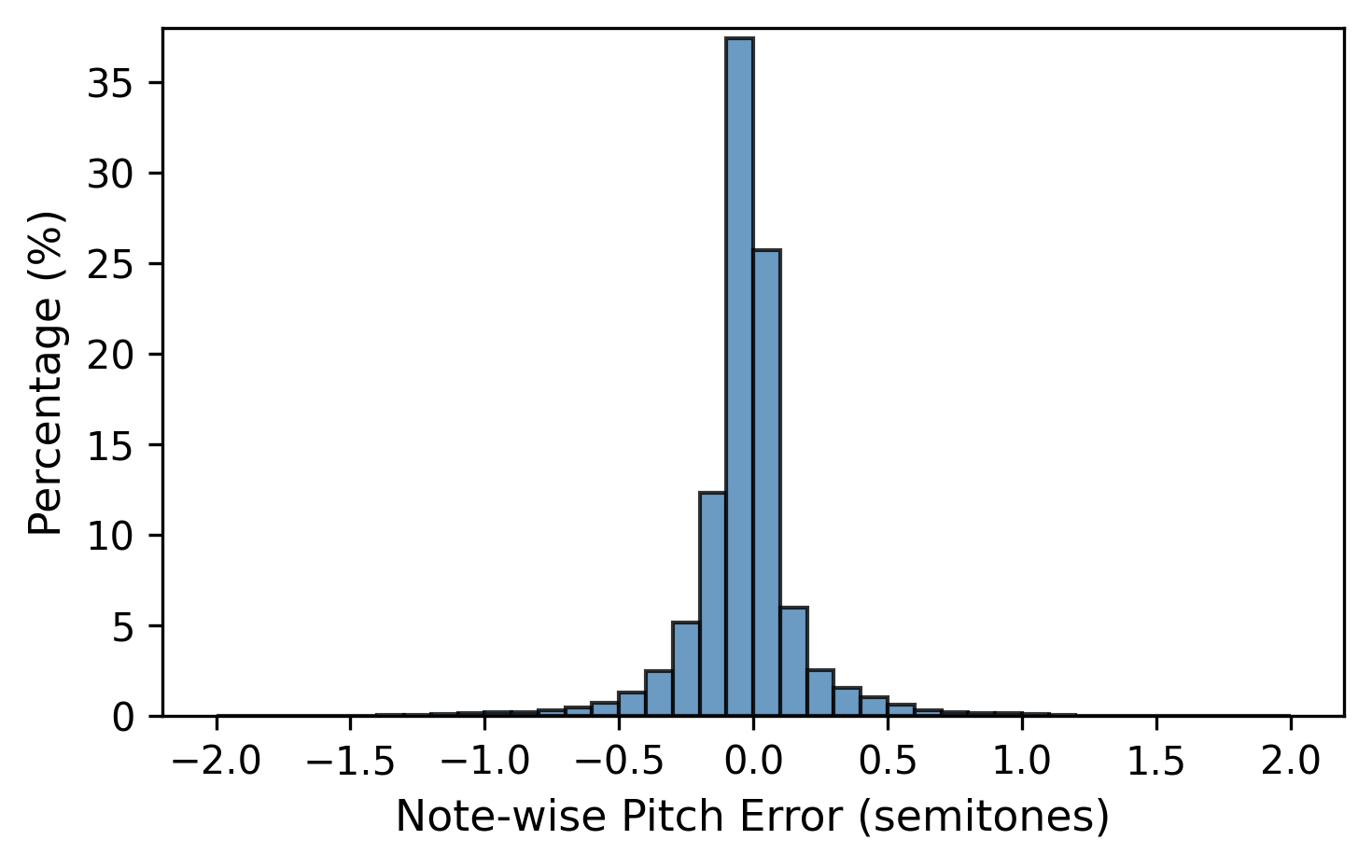}
    \caption*{(a) In-tune subset}
\end{minipage}
\hfill
\begin{minipage}{0.32\textwidth}
    \centering
    \includegraphics[width=\linewidth]{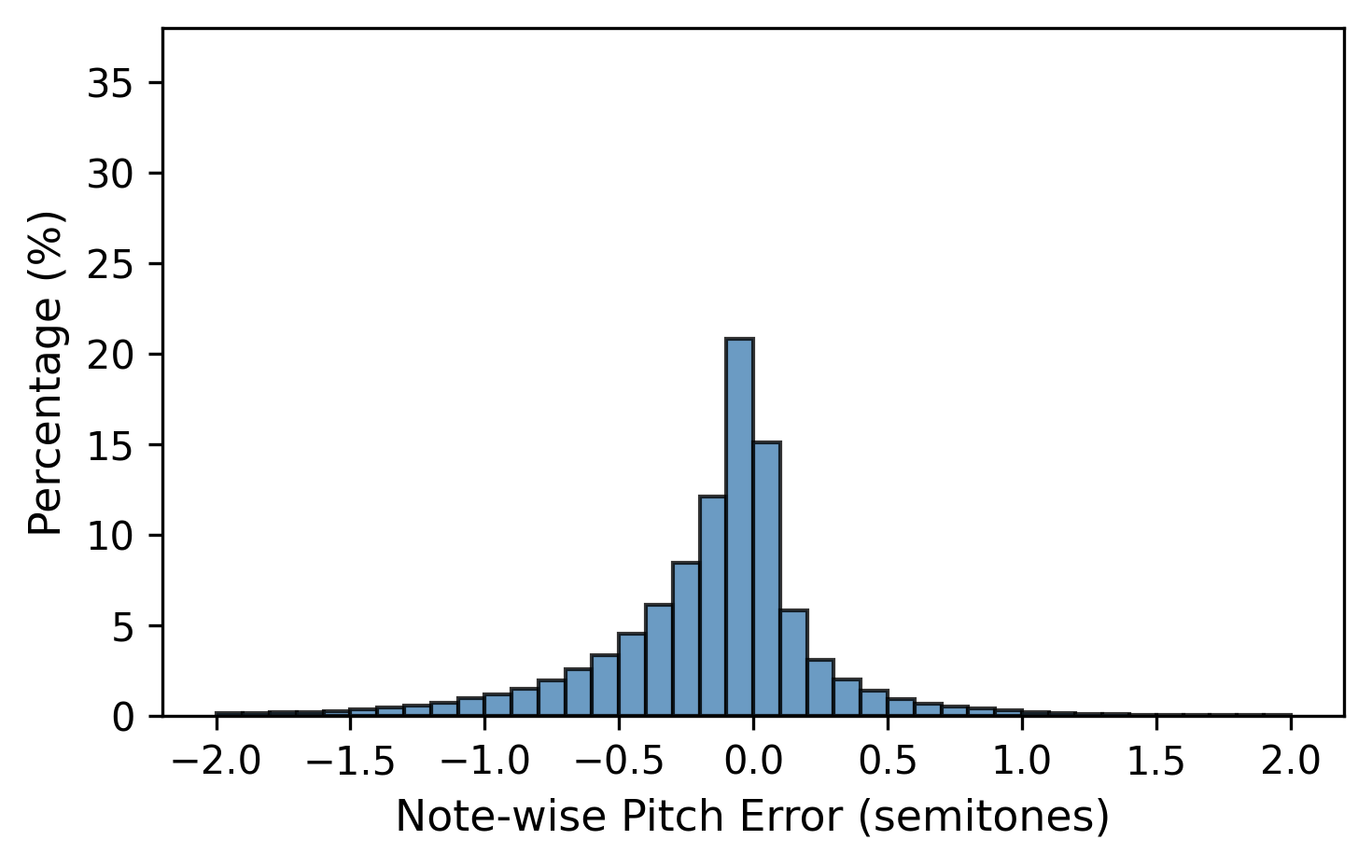}
    \caption*{(b) Moderately detuned subset}
\end{minipage}
\hfill
\begin{minipage}{0.32\textwidth}
    \centering
    \includegraphics[width=\linewidth]{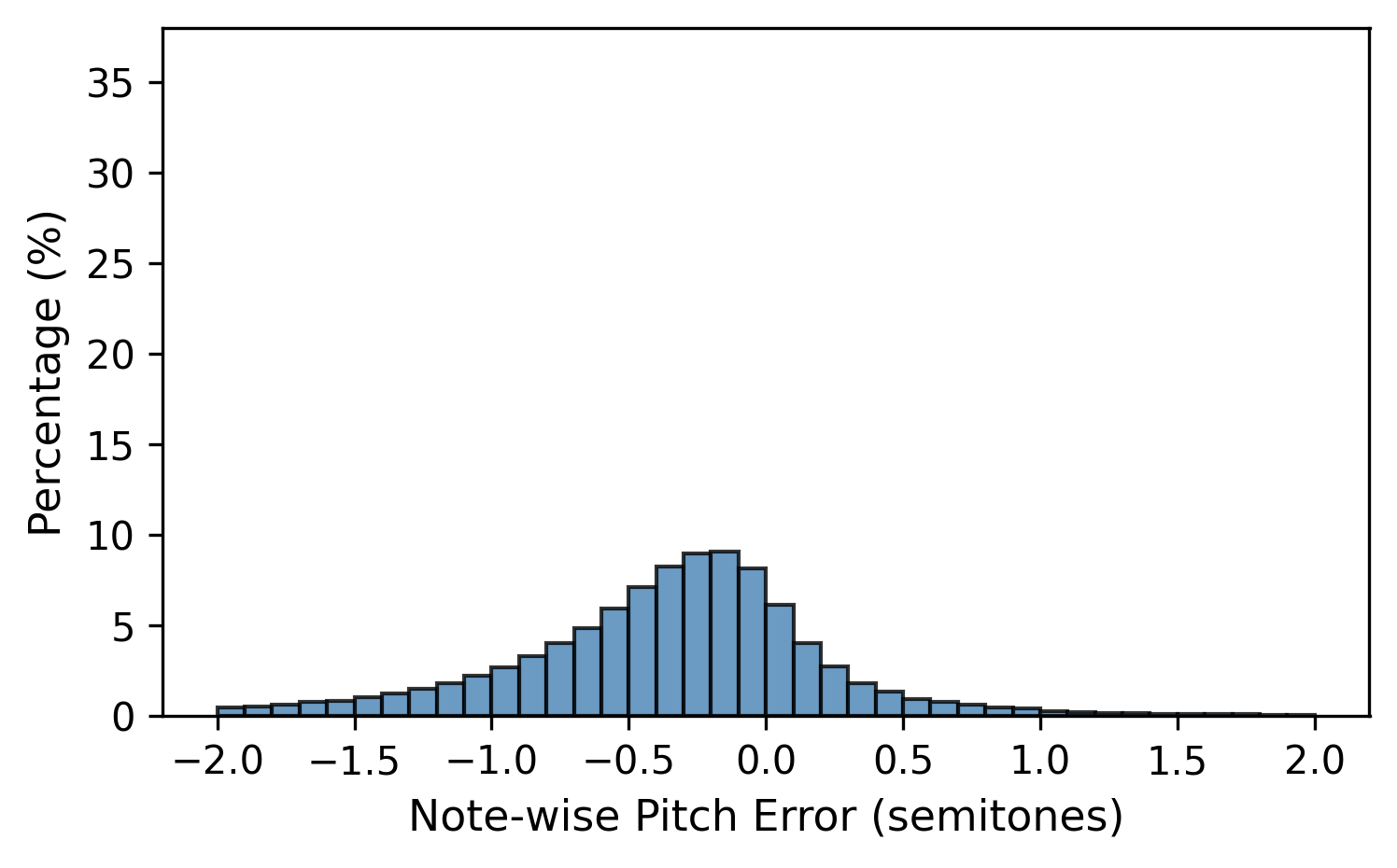}
    \caption*{(c) Highly detuned subset}
\end{minipage}

\caption{
Histograms of note-level pitch errors for the three subsets:
(a) in-tune (10\%), (b) moderately detuned (80\%), and
(c) highly detuned (10\%).  
}
\label{fig:subset_histograms}
\end{figure*}

\begin{table*}[htbp]
\centering
\footnotesize
\begin{tabular}{ll}
\toprule
\textbf{Network} & \textbf{Hyperparameter} \\
\midrule
\textbf{Note Segmentator} \\
\quad Vocal Pitch Projection & Linear(1 $\rightarrow$ 256) \\
\quad Mel-Spectrogram Projection & Linear(80 $\rightarrow$ 256) \\
\quad Feature Fusion & Linear(512 $\rightarrow$ 256) (from concatenated projections) \\
\quad Encoder & Local Transformer (see config below) \\
\quad Output Head & GRU (256 $\rightarrow$ 256) + Linear(256 $\rightarrow$ 1) + Sigmoid \\
\midrule
\textbf{Stationary Pitch Predictor} \\
\quad Vocal Pitch Projection & Linear(1 $\rightarrow$ 256) \\
\quad Mel-Spectrogram Projection & Linear(80 $\rightarrow$ 256) \\
\quad Feature Fusion & Linear(512 $\rightarrow$ 256) (merge concatenated features) \\
\quad Encoder & Local Transformer (see config below) \\
\quad Weight Predictor & Linear(256 $\rightarrow$ 1) \\
\quad Aggregation Function & Softmax over frame-level weights within each note \\
\midrule
\textbf{Local Transformer} \\
\multicolumn{2}{l}{(used in Note Segmentator and Stationary Pitch Predictor)} \\
\quad Layers & 4 \\
\quad Model Dimension & 256 \\
\quad Attention Heads & 4 \\
\quad Head Dimension & 64 \\
\quad Feedforward Block & Linear(256 $\rightarrow$ 1364) + GEGLU + Linear(682 $\rightarrow$ 256) \\ 
\quad Feedforward Expansion Ratio & 4$\times$ (GEGLU-specific, effective ratio $\sim$2.66$\times$) \\ 
\quad Attention Window Size & 512 \\
\quad Residual Paths & 4 parallel streams (split and merge) \\ 
\quad Normalization & Pre-LayerNorm \\ 
\midrule
\textbf{Context-aware Note Pitch Predictor (CNPP)} \\
\quad Transformer Layers & 12 \\
\quad Model Dimension & 768 \\
\quad Feedforward Inner Dimension & 3072 \\
\quad Attention Heads & 12 \\
\quad Activation Function & GELU \\
\quad Dropout / Attention Dropout & 0.1 / 0.1 \\
\quad Positional Encoding & Absolute (learned) \\
\quad Tokenization Granularity & Octuple format (8-token compound events) \\
\quad Downsampling Layer & Linear (6144 $\rightarrow$ 768) (from 8 flattened 768-dim tokens) \\
\quad Upsampling Layer & Linear (768 $\rightarrow$ 6144) (to 8 separate 768-dim tokens) \\
\quad Max Sequence Length & 8192 compound events \\
\midrule
\textbf{Learnable Detuner} \\
\quad Encoder & GRU x2 (64 $\rightarrow$ 64) \\
\quad Output Head & Linear (64 $\rightarrow$ 1) \\
\bottomrule
\end{tabular}
\caption{Architectural hyperparameters of BERT-APC components. Dimensions are denoted as input $\rightarrow$ output.}
\label{tab:apc_hparams}
\end{table*}
\subsection{Dataset and Training Procedure}

In experiments, we utilized a combination of three singing voice datasets: the AI-Hub Guide Vocal Dataset~\cite{dataset_aihub_guidevocal}, the AI-Hub Multi-Singer Singing Dataset~\cite{dataset_aihub_multisinger}, and an in-house collection of diverse vocal recordings. The combined dataset comprises 12,287 samples (509.67 hours).
Each sample consists of a singing voice recording of a song accompanied by a MIDI file containing pitch, duration, lyrics, and other related information. The audio signals were resampled to 22.05 kHz with 16-bit quantization. We extracted vocal pitches using the Praat-Parselmouth library~\cite{parselmouth}, and then converted them to the semitone scale.
Mel-spectrograms were extracted using a hop size of 256, an FFT window size of 1024, a window length of 1024 samples, and 80 Mel bins.

The modules in BERT-APC require training data tailored to their purposes. For example, the stationary pitch predictor requires in-tune samples, whereas the detuner requires highly detuned samples covering a wide range of realistic pitch deviations. 
To obtain these subsets without any pre-trained modules, we first estimate a sample-level pitch error $\bar{\epsilon}$ for each recording by averaging the note-level pitch error $\epsilon_i$ over all notes in the sample. We rank all recordings by $\bar{\epsilon}$ and group the lowest 10\% as in-tune, the middle 80\% as moderately detuned, and the highest 10\% as highly detuned subset.

The distribution of pitch errors across the three subsets is shown in Fig.~\ref{fig:subset_histograms}. For the moderately and highly detuned subsets, 10\% of the data was allocated for validation and another 10\% for testing, with the remaining samples used for training. As a result, the dataset was divided into training (9,828 samples, 406.43 hours), validation (1,229 samples, 44.45 hours), and test sets (1,230 samples, 58.79 hours). The in-tune subset was exclusively used for training the stationary pitch predictor, while the highly detuned subset was used to train the detuner. The CNPP was trained on the moderately detuned subset. The training data for the note segmentator encompasses all training subsets.

\subsection{Implementation Details}
The architectural hyperparameters of all BERT-APC components are summarized in Table~\ref{tab:apc_hparams}. Both the note segmentator encoder $\mathcal{E}_{ns}(\cdot)$ and the stationary pitch predictor encoder $\mathcal{E}_{spp}(\cdot)$ consist of 4-layer Local Attention Transformer~\cite{beltagy2020longformer} with a model dimension of 256, 4 attention heads, and a window size of 512. The segmentator head consists of a single-layer GRU with a sigmoid-activated linear layer, and the stationary pitch predictor head uses a linear layer. Both the note segmentator and stationary pitch predictor were trained with AdamW \cite{adamw17} using a learning rate of $1\text{e}{-5}$, $(\beta_1, \beta_2) = (0.93, 0.98)$, $\epsilon = 1\text{e}{-8}$, and weight decay 0.01. A cosine annealing scheduler was applied with $T_{\text{max}} = 200{,}000$ and $\eta_{\min} = 1\text{e}{-6}$. CNPP is based on MusicBERT-base without architectural modification. We fine-tuned it using AdamW with a learning rate of 1e-5, $(\beta_1,\beta_2) = (0.9, 0.98)$, weight decay of 0.01, and cosine annealing scheduling with $T_{\text{max}} = 500{,}000$ and $\eta_{\min} = 1\text{e}{-6}$. The detuner consists of two layers of GRU with a hidden dimension of 64, followed by a linear pitch error predictor. It was trained with AdamW (batch size 32, $\beta_1{=}0.93$, weight decay 0.01) and a warm-up and cosine annealing schedule ($T_{\text{max}}{=}200{,}000$, $\eta_{\min} = 1\text{e}{-6}$).

We trained all models using Distributed Data Parallel (DDP) on two NVIDIA RTX 3090 GPUs with a total batch size of 32 (16 per GPU).
\subsection{Stationary Pitch Prediction}
To assess the accuracy of note-level stationary pitch estimation, we conducted an evaluation to quantify how closely different pitch estimation methods align with manually annotated stationary pitch. For 6,578 notes, stationary pitch was manually annotated using a custom interface that visualized the vocal pitch contour together with the corresponding note boundaries. For each note, the stable region of the pitch contour was marked, excluding transitional parts such as attacks, releases, or pitch glides. For notes with vibrato, complete and stable vibrato cycles were marked, excluding the initial and final portions where the modulation was not yet stable. The stationary pitch was then computed as the arithmetic mean of the marked region. A piano tone corresponding to the annotated stationary pitch was synthesized and played together with the corresponding singing voice segment for auditory verification. The procedure described above was repeated until the synthesized pitch and the singing voice were perceived as consistent. This annotation protocol follows prior labeling practices in singing voice intonation studies~\cite{manual_pitch_label_cite1, manual_pitch_label_cite2}.

We used two evaluation metrics for stationary pitch estimation: Perceptual Tolerance Rate (PTR) and Mean Absolute Error (MAE).
Previous studies have reported that trained listeners perceive a note as in tune when its pitch deviation falls within a perceptual tolerance range between -10  and +15 cents~\cite{tolerance_range_cite1,tolerance_range_cite2,tolerance_range_cite3,tolerance_range_cite4}, where 1 cent equals 1/100 of a semitone.
Motivated by this observation, we define PTR as the proportion of notes whose pitch estimation error ($\hat{p}_i-p_i^\star$) falls within this perceptual tolerance range, as defined in Eq.~\ref{eq:ptr_def}:
\begin{equation}
    \mathrm{PTR}
    =\frac{100}{N}\sum_{i=1}^{N}\mathbf{1}\big(\hat{p}_i-p_i^\star\in[-10,+15]~\mathrm{cents}\big),
    \label{eq:ptr_def}
\end{equation}
where $\hat{p}_i$ and $p_i^\star$ denote the estimated and human-annotated stationary pitch of the $i$-th note, respectively, and $N$ is the number of evaluated notes. In addition, we also measured the MAE to quantify the average magnitude of the pitch estimation error.


We compared our stationary pitch predictor (SPP) with two existing methods based on note-wise average pitch ~\cite{spp_simple_mean_cite} and weighted median~\cite{svt_phonemesvt23}, respectively. Since this experiment is intended to evaluate the stationary pitch predictor itself, rather than the overall note-pitch prediction performance of a complete SVT model, these note-wise pitch estimation methods provide more appropriate comparison models in this setting. The quantitative results are summarized in Table~\ref{tab:ptr_mae}. We evaluated them using the human annotated test samples. The first baseline method averages all frame-level pitches within a note interval. This method achieved the lowest accuracy, exhibiting a PTR of 76.9\% and an MAE of 10.9~cents. Especially, it was highly susceptible to transitional regions such as attacks, releases, and strong vibrato, as shown in Fig.~\ref{fig:pitch_comparison}(a).
The second baseline method, which is based on a weighted median rule, corresponds to the pitch estimation strategy used in PhonemeSVT~\cite{svt_phonemesvt23}.
It achieved better performance, with a PTR of 89.2\% and an MAE of 5.7~cents. By placing higher weights on frames near the note center, this method demonstrated reduced susceptibility to transitional regions. Nevertheless, it yielded substantial estimation errors for notes exhibiting highly asymmetric transitions or off-center stationary regions, as illustrated in Fig.~\ref{fig:pitch_comparison}(b).

In contrast, the proposed SPP achieved the highest PTR (94.3\%) and the lowest MAE (3.5~cents) among all methods.  
As shown in Fig.~\ref{fig:pitch_comparison}, SPP effectively suppressed transitional regions by assigning high weights to stationary regions without making any assumptions about their positions, and it successfully detected off-center stationary regions.
Our results suggest that the data-driven stationarity-weight estimation approach detects stationary regions more effectively and is more robust to pitch variation than simple statistical methods.

\begin{table}[t]
\centering
\caption{Stationary pitch prediction performance, evaluated against manually annotated stationary pitches.}
\label{tab:ptr_mae}
\begin{tabular}{lcc}
\toprule
\textbf{Method} & \textbf{PTR (\%)} $\uparrow$ & \textbf{MAE (cents)} $\downarrow$\\
\midrule
Average & 76.9 & 10.9 \\
Weighted Median & 89.2 & 5.7 \\
\textbf{SPP (Ours)} & \textbf{94.3} & \textbf{3.5} \\
\bottomrule
\end{tabular}
\end{table}

\subsection{Note Pitch Prediction}

\begin{table}[t]
\centering
\small
\setlength{\tabcolsep}{5pt}
\renewcommand{\arraystretch}{1.2}
\begin{tabular}{l|c|c}
\toprule
\textbf{\makecell[l]{Note Pitch \\ Predictor}} & \makecell{\textbf{RPA (\%) $\uparrow$} \\ \textbf{(Moderately Detuned)}} & \makecell{\textbf{RPA (\%) $\uparrow$} \\ \textbf{(Highly Detuned)}} \\
\midrule
Frame-level Rounding & 85.07 & 58.02 \\
PhonemeSVT & 81.36 & 55.65 \\
ROSVOT         & 89.60 & 78.75 \\
\textbf{\makecell[l]{BERT-APC\\(Ours)}} & \textbf{94.95} & \textbf{89.24} \\
\midrule
\bottomrule
\end{tabular}
\caption{Note pitch prediction performance in terms of RPA on the moderately and highly detuned test subsets.}
\label{tab:frame_pitch_metrics}
\end{table}

We employ the Raw Pitch Accuracy (RPA) metric to quantitatively evaluate the accuracy of target note pitch prediction, which measures the proportion of voiced frames whose predicted note pitch lies within 0.5 semitones of the GT note pitch.
Although RPA is computed from frame-level pitches, it has also been used in recent SVT studies~\cite{svt_rosvot24, svt_stars25}, to evaluate note-level pitch prediction performance. Following those studies, we adopt RPA to assess pitch prediction performance at the note level.
\begin{equation}
\label{eq:RPA}
\text{RPA} = \frac{1}{|\mathcal{V}|} \sum_{t \in \mathcal{V}} \mathbf{1}\left( |\hat{q}^f_t - q^f_t| < 0.5 \right),
\end{equation}
where $\mathcal{V}$ is the set of voiced frames, while $\hat{q}^f_t$ and $q^f_t$ are the predicted and the GT note pitches converted to the frame-level resolution, respectively. 

To convert note-level predictions to the frame-level resolution, we repeat each predicted note pitch over its predicted note duration and each GT note pitch over the corresponding GT note duration, thereby obtaining predicted and ground-truth frame-level note-pitch sequences on the frame grid. When RPA is used as a note-level metric, as in prior SVT studies~\cite{svt_rosvot24, svt_stars25} and the present study, it effectively represents a duration-weighted average of note-level threshold-based accuracy, which reflects the proportionally larger impact of longer notes on the pitch correction process in Stage 3.


We compared BERT-APC with two recent SVT models: PhonemeSVT~\cite{svt_phonemesvt23}, which predicts note boundaries and estimates note pitch using the weighted median of frame-level pitch values within the predicted note boundaries, and ROSVOT~\cite{svt_rosvot24}, which employs a deep learning-based pitch classifier to predict discrete note pitches. We selected these two models because they represent two distinct design paradigms---statistical aggregation and discrete pitch classification---and provide publicly available open-source implementations. Furthermore, we added a simple closest-discrete-pitch rounding baseline, which corresponds to the straightforward solution of mapping the input vocal pitch to the closest discrete pitch without leveraging musical context. Specifically, this rounding is performed at the frame level: each frame's vocal pitch is independently quantized to the nearest semitone, without note-level aggregation. It thus provides a learning-free reference against which the learned note-level approaches are compared. This baseline is conceptually related to Auto-Tune, which applies pitch quantization without note-level aggregation. However, it is not identical to Auto-Tune, whose detailed algorithm is not fully disclosed.


Table~\ref{tab:frame_pitch_metrics} presents the evaluation results on the moderately and highly detuned test subsets. Compared with the closest-discrete-pitch rounding baseline, ROSVOT achieves higher RPA by 4.53 pp on the moderately detuned subset and by 20.73 pp on the highly detuned subset, suggesting that learned note-pitch prediction is more effective than simple pitch quantization for this task. BERT-APC further improves over ROSVOT by 5.35 pp and 10.49 pp on the moderately and highly detuned subsets, respectively, suggesting the benefit of leveraging musical context in target note-pitch prediction.

Another notable trend is that the performance drops from the moderately to the highly detuned subset are much larger for the closest-discrete-pitch rounding baseline and PhonemeSVT than for ROSVOT and BERT-APC. Specifically, the RPA decreases by 27.05 pp for the rounding baseline and by 25.71 pp for PhonemeSVT, whereas the drops are 10.85 pp for ROSVOT and 5.71 pp for BERT-APC. This observation suggests that methods based more directly on the observed pitch values---through direct quantization or statistical aggregation---are more affected when the input pitch deviates substantially from the GT note pitch under the $\pm 0.5$-semitone RPA criterion.

Interestingly, the closest-discrete-pitch rounding baseline yields slightly higher RPA than PhonemeSVT, by 3.71 pp on the moderately detuned subset and by 2.37 pp on the highly detuned subset. Because PhonemeSVT predicts one representative pitch per note and expands it to the frame level for RPA computation, representative-pitch or note-interval errors can reduce the score over many frames at once. By contrast, frame-level rounding evaluates each frame independently, so some frames within a note can still satisfy the $\pm 0.5$-semitone criterion even when the note is not matched consistently.

To examine how the predicted target note pitches are realized in the corrected singing voice, we additionally measured the RPA between the GT note pitch and the pitch extracted from the corrected singing voice produced by BERT-APC. The RPA values were 89.90\% and 84.20\% on the moderately and highly detuned subsets, respectively. Although these values are lower than the RPA values in Table~\ref{tab:frame_pitch_metrics}, they still indicate that the corrected singing voice follows the GT note pitch with reasonably high accuracy. We observed that the lower RPA mainly arises from expressive pitch variations in the corrected singing voice, such as vibrato and transitions, which make some frames deviate from the GT note pitch by more than 0.5 semitones.


\begin{table}[t]
\centering
\setlength{\tabcolsep}{5pt}
\renewcommand{\arraystretch}{1.2}
\begin{tabular}{l|c|c}
\toprule
\textbf{Model} & \textbf{\makecell{Pitch\\Accuracy}} & \textbf{\makecell{Expression\\Preservation}} \\
\midrule
Auto-Tune & $3.22 \pm 0.18$ & $3.81 \pm 0.17$ \\
Melodyne & $3.08 \pm 0.18$ & $\mathbf{3.85 \pm 0.17}$ \\
\textbf{BERT-APC (Ours)} & $\mathbf{4.32 \pm 0.15}$ & $3.80 \pm 0.17$ \\
\bottomrule
\end{tabular}
\caption{The MOS evaluation results on a 5-point Likert scale are presented as the mean and 95\% confidence interval, averaged across all participants and test samples.}
\label{tab:mos}
\end{table}

\begin{figure*}[!t]
    \centering
    \subfloat[Auto-Tune]{%
        \includegraphics[width=0.31\linewidth]{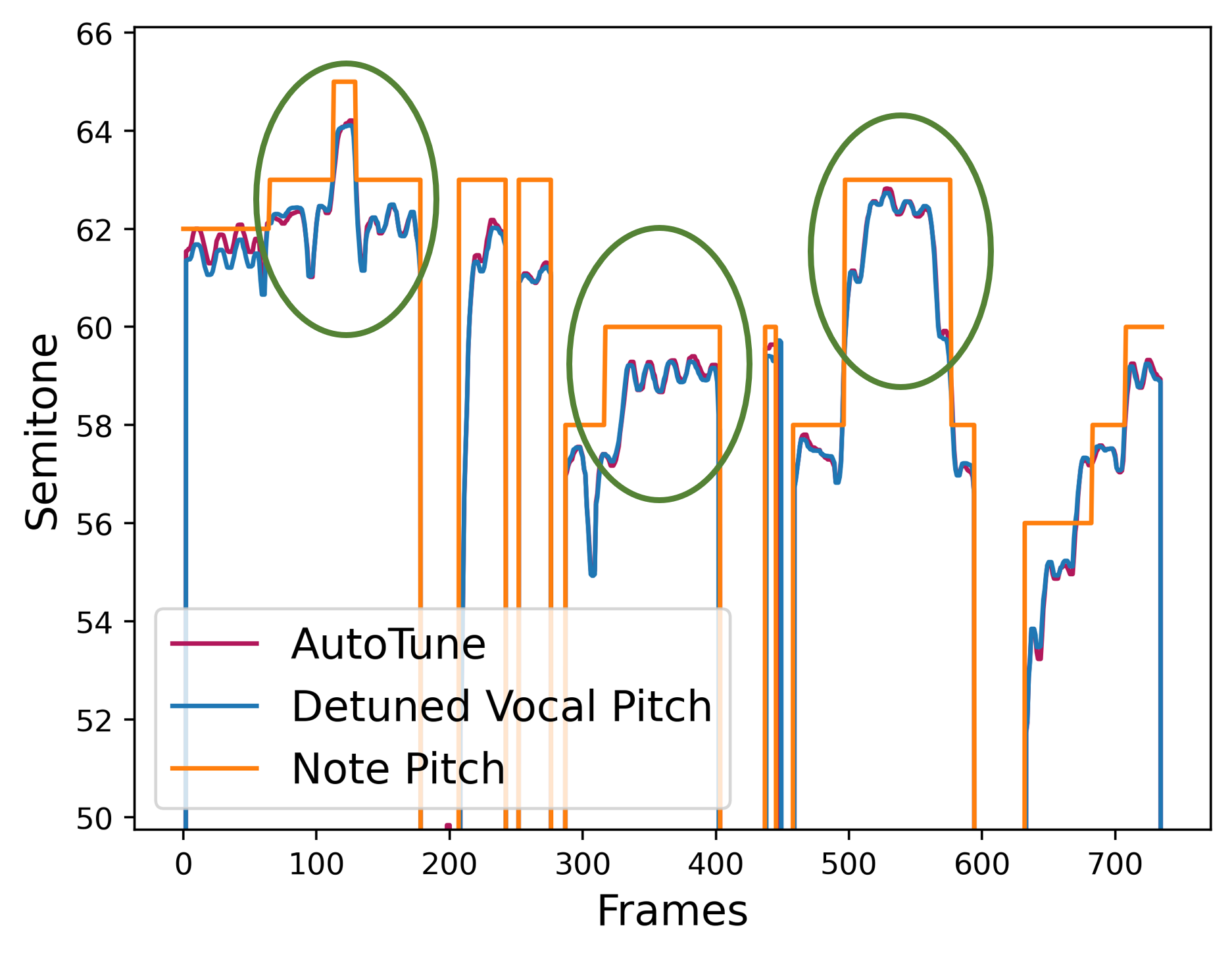}%
        \label{fig:correction_autotune}}
    \hfil
    \subfloat[Melodyne]{%
        \includegraphics[width=0.31\linewidth]{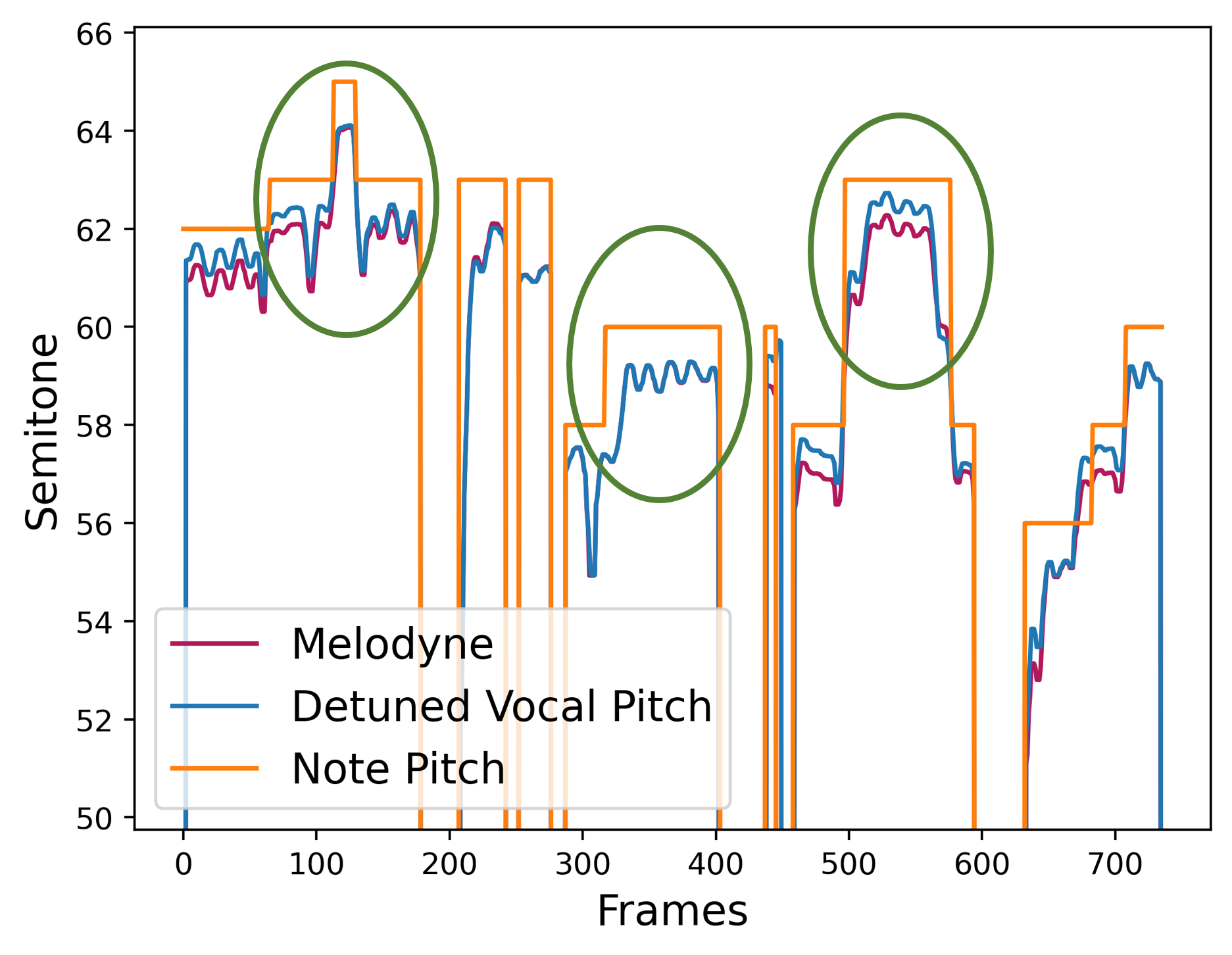}%
        \label{fig:correction_melodyne}}
    \hfil
    \subfloat[BERT-APC (Ours)]{%
        \includegraphics[width=0.31\linewidth]{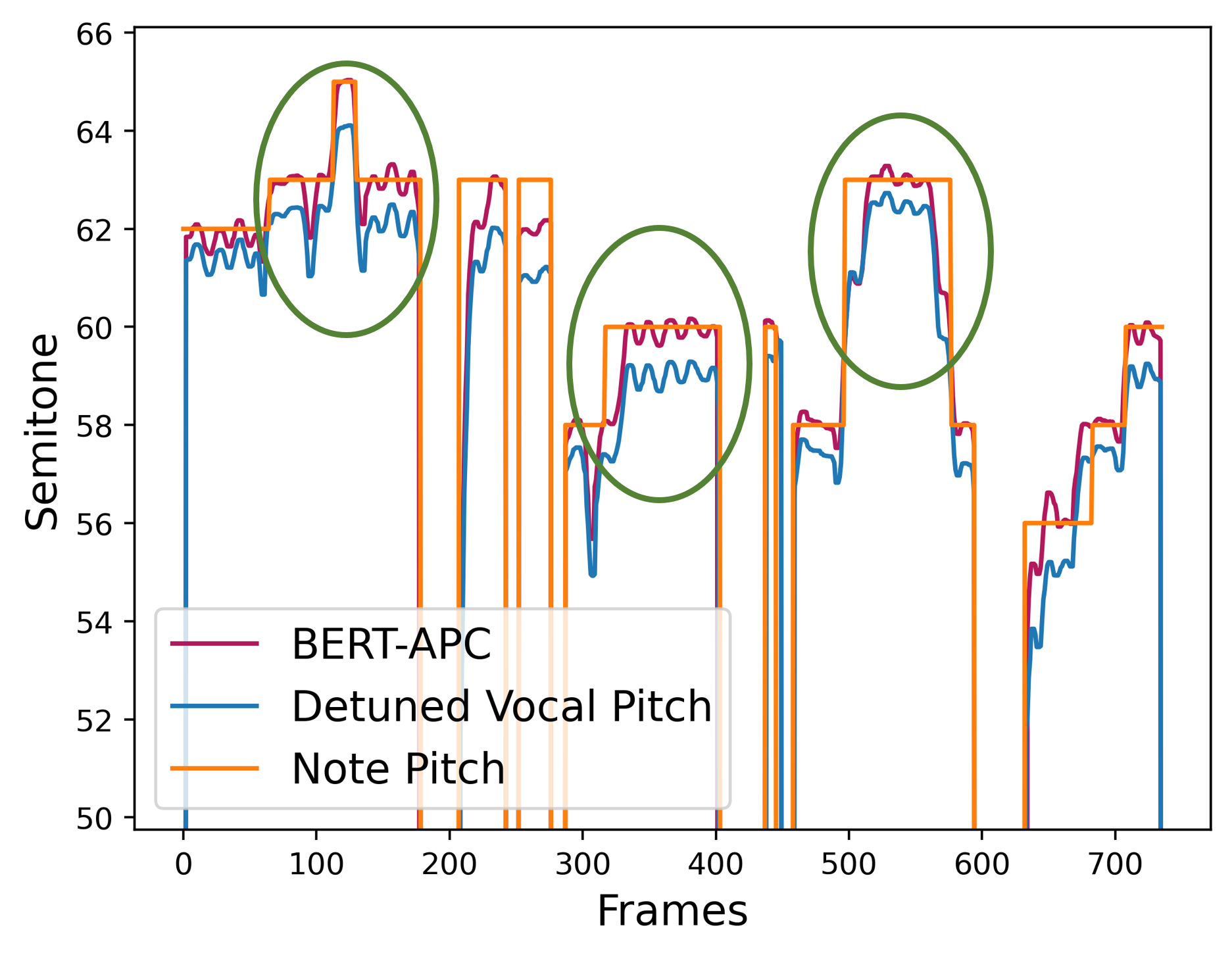}%
        \label{fig:correction_ours}}
    \caption{Visualization of pitch correction results for a highly detuned sample. 
    The green, blue, and orange lines represent the correction results, the input pitch, and the GT note pitch, respectively. 
    (a) Auto-Tune and (b) Melodyne failed to correct pitch deviations exceeding one semitone, especially when the deviation spanned the full pitch range of a note. 
    (c) In contrast, BERT-APC successfully corrected them by leveraging musical context via the musical language model, MusicBERT.
    }
    \label{fig:correction_comparison}
\end{figure*}

\begin{figure*}[t]
\centering
\begin{minipage}[t]{0.32\textwidth}
  \centering
  \includegraphics[width=\linewidth]{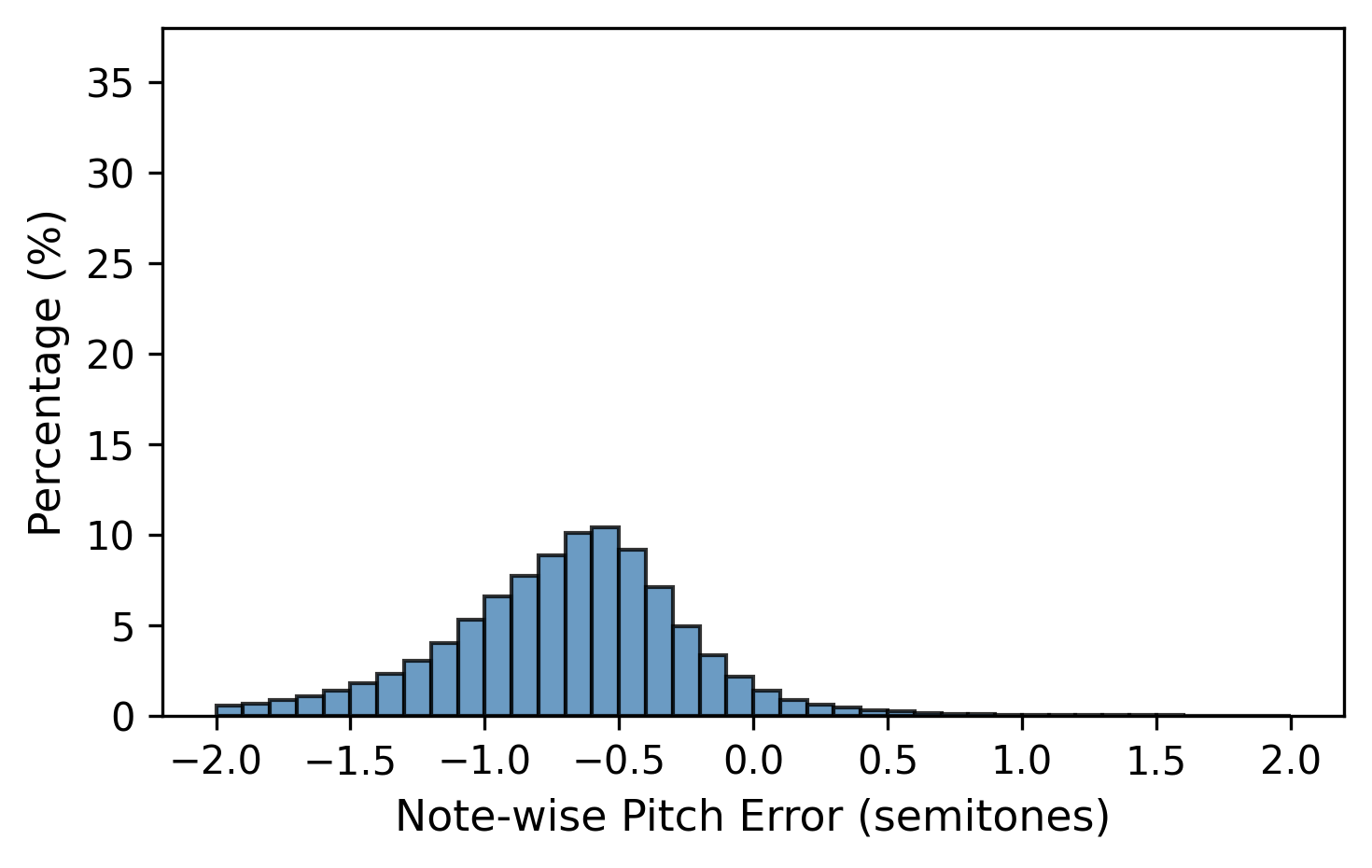}
  \caption*{(a) Detuned by the learnable detuner}
  \label{fig:moderately_detuned_reference}
\end{minipage}
\hfill
\begin{minipage}[t]{0.32\textwidth}
  \centering
  \includegraphics[width=\linewidth]{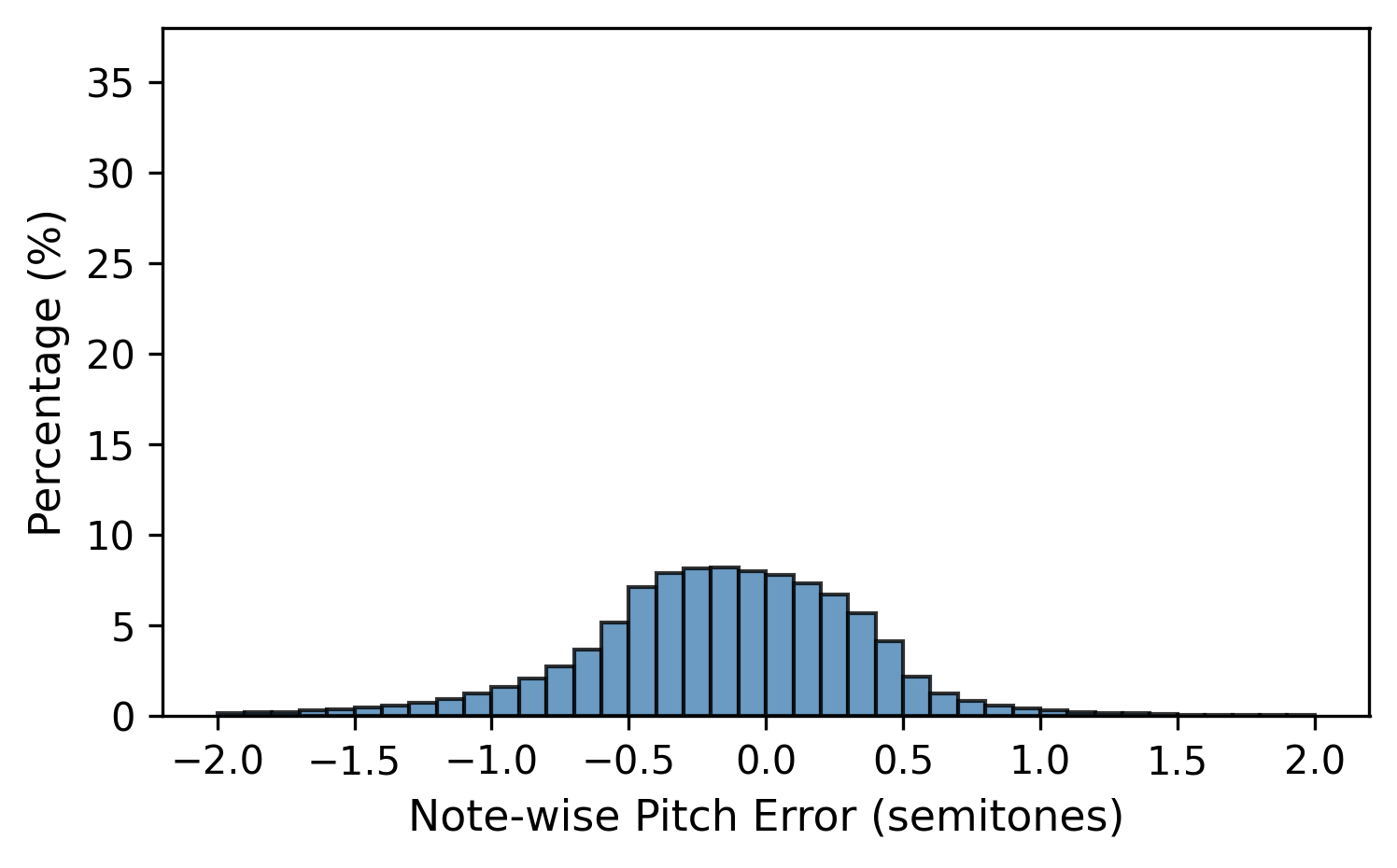}
  \caption*{\centering
    (b) Randomly detuned $\mathrm{Uniform}(-0.5, +0.5)$}
  \label{fig:learnable_detune}
\end{minipage}
\hfill
\begin{minipage}[t]{0.32\textwidth}
  \centering
  \includegraphics[width=\linewidth]{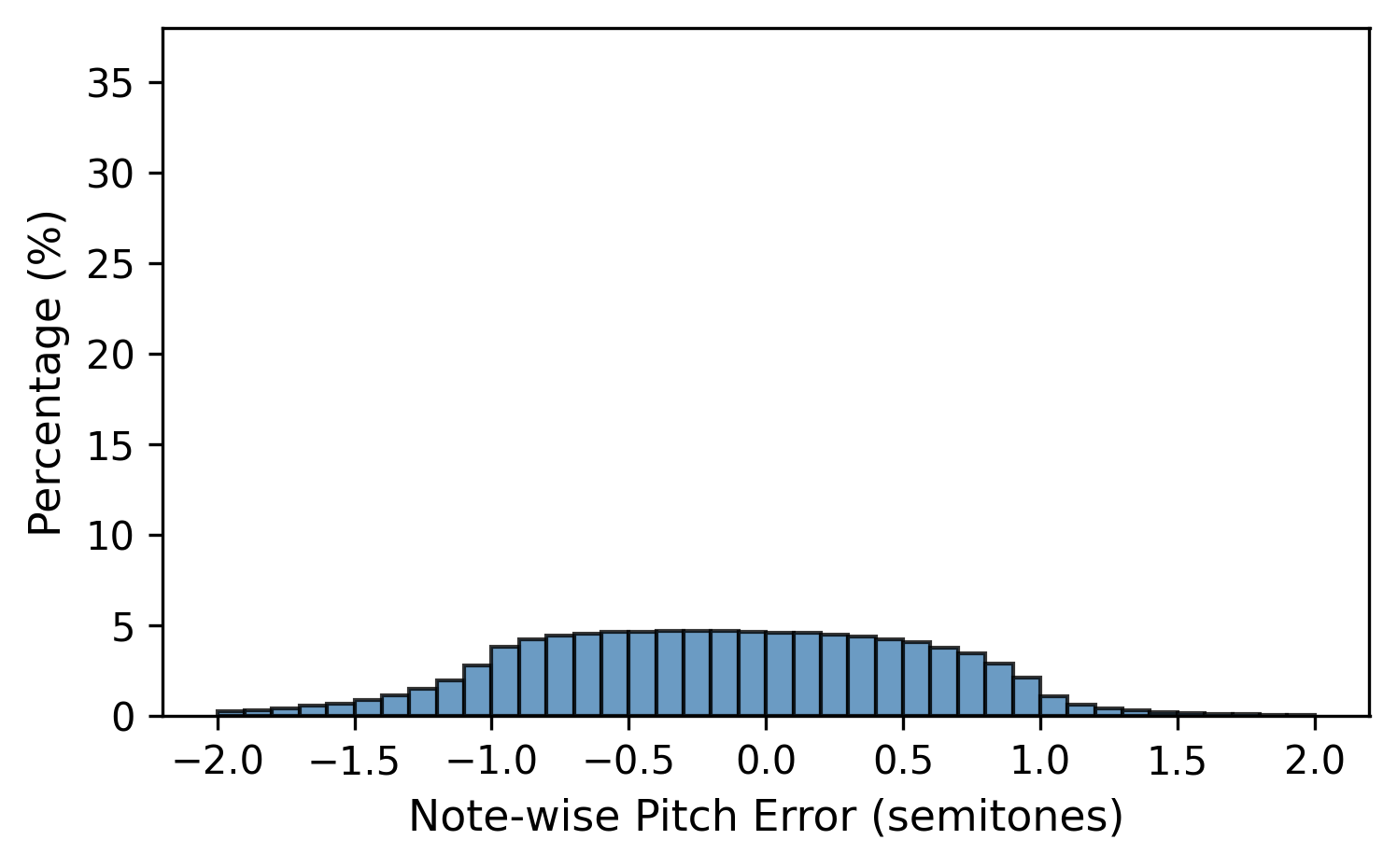}
  \caption*{\centering
  (c) Randomly detuned $\mathrm{Uniform}(-1, +1)$}
  \label{fig:random_detune}
\end{minipage}

\caption{
The distribution of note-level pitch errors for the moderately detuned dataset under three augmentation strategies. The proposed learnable detuner produces a distribution resembling the highly detuned subset (Fig.~\ref{fig:subset_histograms}(c)), whereas random detuning yields substantially different distributions.
}
\label{fig:detuner_comparison}
\end{figure*}

\subsection{MOS Tests}
We assessed the perceptual quality of the models in pitch correction accuracy and expression-preservation ability via Mean Opinion Score (MOS) tests. For the MOS test, we collected Korean singing recordings from three amateur singers (two male and one female) and divided them into excerpts of up to 10 seconds. These excerpts showed an average pitch error of -0.57 semitones with a standard deviation of 0.38, suggesting that the excerpts were noticeably detuned overall and tended to be flat. In addition, the excerpts generally contained expressive singing characteristics, such as vibrato and natural pitch transitions between notes.

34 listeners participated in the MOS test, and each listener rated 24 samples. Of these, 12 had formal musical training and 22 were general listeners without formal training. The participants included 16 males and 18 females, with age groups of 20--29 ($n=7$), 30--39 ($n=10$), 40--49 ($n=5$), and 50+ ($n=12$). All participants provided informed consent prior to the test.

The participants evaluated pitch accuracy and expression-preservation of pitch-corrected singing voice, using a 5-point Likert scale. Before the test, participants were given scoring guidelines for both tasks, along with low and high anchor reference audio. In the pitch accuracy task, each corrected sample was presented alongside the corresponding GT note pitches rendered with a piano sound to help participants identify pitch deviations more easily. In the expression-preservation task, each corrected sample was presented together with the original singing voice, and participants evaluated how well the expressive elements of the original singing were preserved. The piano reference was omitted to avoid bias from perceived pitch deviations, and participants were instructed to ignore pitch accuracy and audio quality.


For comparison, we used two widely used commercial tools as baseline models: Melodyne and Auto-Tune. To apply pitch correction with Auto-Tune, the musical key of each sample was first identified using Auto-Key, followed by correction with Auto-Tune Pro. The following parameters were applied uniformly across all samples:

\begin{itemize}
    \item \textbf{Retune Speed (30)}: Determines how quickly pitch deviations are corrected. A moderate value allows natural transitions while minimizing pitch errors.
    \item \textbf{Flex Tune (50)}: Balances correction with the singer’s intended pitch deviations, preserving expressive slides and bends.
    \item \textbf{Natural Vibrato (0)}: Maintains the original vibrato without artificial modification.
    \item \textbf{Humanize (50)}: Reduces robotic artifacts by softening pitch correction on sustained notes.
\end{itemize}

For Melodyne, the following settings were used:

\begin{itemize}
    \item \textbf{Pitch Center (100)}: Aligns each note exactly to the target pitch, minimizing cent-level deviations.
\end{itemize}

All parameter settings were determined in consultation with professional audio engineers to ensure natural-sounding pitch correction while preserving expressive qualities in the vocal performance.


Table~\ref{tab:mos} presents the results of the MOS test. BERT-APC notably outperformed the baseline methods in pitch accuracy, achieving the highest score of $4.32 \pm 0.15$. This score surpasses that of the baseline models, Auto-Tune ($3.22 \pm 0.18$) and Melodyne ($3.08 \pm 0.18$), by substantial margins.
To examine whether the observed differences in the MOS for pitch accuracy among systems were statistically significant, we conducted a one-way analysis of variance (ANOVA), followed by post-hoc pairwise comparisons with Bonferroni correction. The ANOVA resulted in $F(2, 405) = 60.46$, $p < 0.001$, $\eta^{2} = 0.23$, indicating that the systems differed in their mean MOS for pitch accuracy.
Furthermore, the post-hoc tests yielded $p$-values below 0.001 in the pairwise comparisons between BERT-APC and Auto-Tune, and between BERT-APC and Melodyne.
These results indicate that the differences in MOS for pitch accuracy between BERT-APC and each baseline are statistically significant.

Regarding expression-preservation ability, the commercial APC, Melodyne achieved the highest score of ($3.85 \pm 0.17$), but Auto-Tune ($3.81 \pm 0.17$) and BERT-APC ($3.80 \pm 0.17$) also demonstrated comparable scores. In summary, BERT-APC yielded significantly higher MOS scores in pitch accuracy while maintaining expression preservation comparable to Auto-Tune and Melodyne. These results indicate that contextual information plays an important role in target note pitch estimation from detuned singing voices.

Fig.~\ref{fig:correction_comparison} displays the correction results of each model for a highly detuned sample. Auto-Tune and Melodyne were unable to correct high pitch deviations exceeding one semitone, particularly when these deviations spanned the entire duration of the note, as highlighted by the green ellipses in Fig.~\ref{fig:correction_comparison}(a) and (b). In contrast, as shown in Fig.~\ref{fig:correction_comparison}(c), BERT-APC demonstrated notably improved correction performance. 

\begin{figure*}[ht!]
    \centering
    \includegraphics[width=0.8\textwidth]{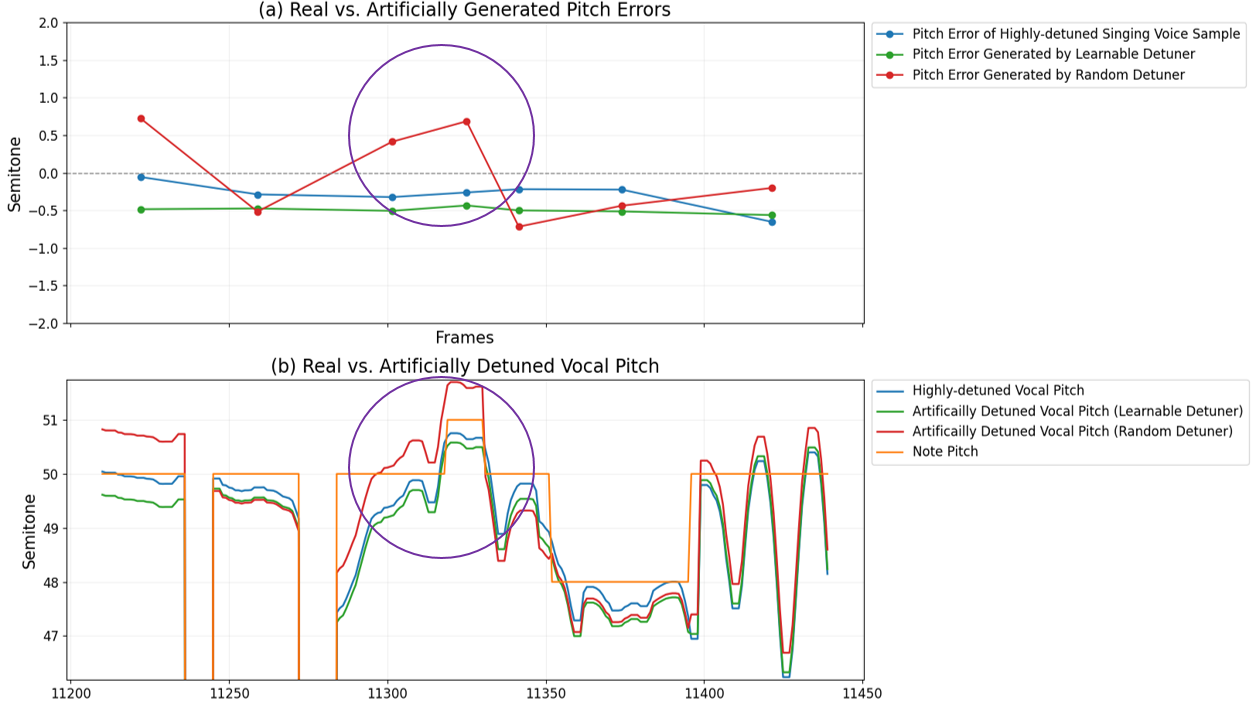}
    \caption{(a) Note-level pitch error sequences synthesized by the learnable and random detuners, compared with that of a highly detuned singing voice sample. 
    (b) The detuned pitch contours produced by adding the synthesized pitch errors to an in-tune baseline pitch contour, which was obtained by removing note-level pitch errors from a highly detuned vocal pitch. The learnable detuner more closely imitates the temporal pattern of the real detuned sample, whereas the random detuner produces less consistent note-to-note fluctuations.}
    \label{fig:detuner_temporal_pattern}
\end{figure*}

\subsection{Analysis of Generated Pitch Error}

To evaluate the effectiveness of our learnable detuner, we visualized the distribution of the pitch errors it produced. For comparison, we implemented two baseline data augmentation algorithms that generate pitch errors from uniform distributions, $Uniform(-0.5,+0.5)$ and $Uniform(-1,+1)$, following \cite{apc_deepautotuner20}. In Fig.~\ref{fig:detuner_comparison}(a), the proposed learnable detuner produced a pitch error distribution for the moderately detuned samples that was closely similar to that of the highly detuned subset (Fig.~\ref{fig:subset_histograms}(c)), indicating its effectiveness in reproducing real-world detuning patterns. In contrast, pitch errors generated by the random detune algorithm differ substantially from this distribution, as shown in Fig.~\ref{fig:detuner_comparison}(b) and (c).

To quantify this observation, we additionally measured the first-order Wasserstein distance (WD) between each generated distribution and that of the highly detuned subset. The proposed learnable detuner achieved a lower WD ($0.363$) than $Uniform(-0.5,+0.5)$ ($0.453$) and $Uniform(-1,+1)$ ($0.446$), indicating that it more closely imitates the pitch error distribution of the highly detuned subset.

In real singing, pitch errors in adjacent notes often exhibit temporal dependency, with similar error tendencies persisting across multiple consecutive notes, as illustrated in Fig.~\ref{fig:detuner_temporal_pattern}. The learnable detuner is trained to imitate this note-to-note dependency by generating each note-level pitch error autoregressively conditioned on previously generated errors. To evaluate whether the model captures such temporal dependencies, we compute the first- and second-order differences—defined as ($p_t - p_{t-1}$) and ($p_{t+1} - 2p_t + p_{t-1}$), respectively—of the synthesized pitch error sequences. The first-order difference represents the local slope, whereas the second-order difference represents the local curvature. We then measure their MAEs against those computed from pitch error sequences of the highly detuned samples.

We also report the MAE between pitch errors of artificially detuned samples and those of highly detuned samples to assess the similarity of note-level error distributions. Table~\ref{tab:detuner_temporal_metrics} summarizes the results. The learnable detuner achieves substantially lower MAEs, indicating that the proposed method effectively reproduces the error patterns of real detuned samples.

\begin{table}[h]
\centering
\caption{MAEs of note-level pitch errors, first-order differences representing local slopes, and second-order differences representing local curvatures, computed between real and synthesized pitch error sequences.}
\label{tab:detuner_temporal_metrics}
\small
\setlength{\tabcolsep}{4.5pt}
\renewcommand{\arraystretch}{1.1}
\begin{tabular}{lccc}
\toprule
Detuning Methods
& \shortstack[c]{Pitch Error\\MAE ($\downarrow$)}
& \shortstack[c]{1st-order Diff.\\MAE ($\downarrow$)}
& \shortstack[c]{2nd-order Diff.\\MAE ($\downarrow$)} \\
\midrule
Random Detuner\\
\hspace{1em}$U(-1.0,+1.0)$ & 0.63 & 0.82 & 1.40 \\
\hspace{1em}$U(-0.5,+0.5)$ & 0.45 & 0.57 & 0.96 \\
\textbf{Learnable Detuner} & \textbf{0.24} & \textbf{0.40} & \textbf{0.72} \\
\bottomrule
\end{tabular}
\end{table}



\subsection{Ablation Studies}
To assess the contribution of the proposed methods, we conducted ablation studies on three key components:
(1) the data augmentation with the learnable detuner,
(2) the interpolated pitch embedding representation, and
(3) the context-aware note pitch predictor (CNPP).
We constructed three variants of BERT-APC, each disabling or modifying a specific component while leaving the rest of the architecture unchanged. We measured the RPA of the models, and the results are summarized in Table~\ref{tab:ablation_results}.
\begin{table}[h]
\centering
\small
\begin{tabular}{l|c|c}
\toprule
\textbf{Models} & \textbf{\makecell{Moderately \\ Detuned (\%)}} & \textbf{\makecell{Highly \\ Detuned (\%)}} \\
\midrule
BERT-APC & \textbf{94.95} & \textbf{89.24} \\
w/o data augmentation & \underline{94.94} & 87.41 \\
w/o interp. pitch embedding & 94.74 & \underline{88.18} \\
w/o CNPP & {94.12} & {71.25} \\
\bottomrule
\end{tabular}
\caption{The results of ablation studies.}
\label{tab:ablation_results}
\end{table}

For the moderately detuned subset, the performance differences among the four models were relatively small, with only a 0.83 pp gap between BERT-APC and the lowest-performing model, `w/o CNPP'.
However, for the highly detuned subset, disabling each component resulted in larger performance drops. When data augmentation was removed, the performance decreased by 1.83 pp, suggesting that the proposed augmentation provides additional robustness under severe detuning conditions. Nevertheless, the `w/o data augmentation' model still exhibits competitive performance, suggesting that, being built upon a pretrained MusicBERT, CNPP has already acquired rich knowledge for pitch prediction from large-scale symbolic music data and is therefore less prone to overfitting.

When interpolated pitch embedding was replaced with the embedding of the closest discrete pitch, the performance  dropped by 1.06 pp, which is moderate. We attribute this to the ability of CNPP to recover the correct discrete pitches by leveraging contextual information, even when some notes are incorrectly mapped to discrete pitches due to simple rounding.

In contrast, replacing CNPP with a simple note-level rounding algorithm that selects the closest discrete pitch to the stationary pitch resulted in the most significant performance drop of 17.99 pp. These results indicate that the proposed methods are effective in improving robustness, showing greater improvements for highly detuned samples. In particular, CNPP was highly effective in correcting severely detuned pitch sequences.



Since BERT-APC explicitly produces a frame-level target F0 contour, it can be combined with any pitch adjustment module. We built another ablation model by replacing Praat-Parselmouth with a modern pitch-controllable neural vocoder, SiFiGAN~\cite{vocoder_sifigan23}.

Table~\ref{tab:pitch_adjustment_obj} presents the evaluation results. Praat-Parselmouth achieves better pitch control accuracy than SiFiGAN, exhibiting lower F0 RMSE and V/UV error rate. However, the gap is modest, indicating that SiFiGAN is a feasible candidate for the pitch adjustment module of BERT-APC. Moreover, while Praat-Parselmouth introduces audible artifacts in some samples, SiFiGAN produces cleaner audio in those cases. Representative audio examples are available on the demo page. 

\begin{table}[t]
\centering
\caption{Comparison of pitch adjustment modules. $F_0$ RMSE is computed in Hz over voiced frames only.}
\label{tab:pitch_adjustment_obj}
\resizebox{\columnwidth}{!}{
\begin{tabular}{lcc}
\toprule
\textbf{Pitch Adjustment Modules} & F0 RMSE (Hz) $\downarrow$ & \textbf{V/UV error (\%) $\downarrow$} \\
\midrule
Praat-Parselmouth (TD-PSOLA) & 9.79 & 1.02 \\
SiFiGAN & 10.37 & 1.37 \\
\bottomrule
\end{tabular}%
}
\end{table}

\subsection{Sensitivity to Note Segmentation Errors}
Since the downstream components of BERT-APC rely on the predicted note boundaries, we conducted an additional experiment to evaluate how note segmentation errors affect downstream note-pitch prediction performance. Specifically, we randomly perturbed each predicted note boundary by up to $\pm 20$, $\pm 40$, $\pm 80$, or $\pm 160$ ms, while only accepting perturbations for which both neighboring notes remained at least two frames long ($\approx 23$ ms).

As shown in Table~\ref{tab:boundary_perturbation}, BERT-APC maintained stable performance under moderate boundary perturbations, with RPA decreases of 0.50 pp on both the moderately and highly detuned subsets at $\pm 20$ ms, 1.40 pp and 1.30 pp at $\pm 40$ ms, and 3.80 pp and 3.20 pp at $\pm 80$ ms, respectively. Under the larger perturbation of $\pm 160$ ms, the degradation became more noticeable, with RPA drops of 9.10 pp and 7.50 pp on the moderately and highly detuned subsets, respectively. These results indicate that BERT-APC is not overly sensitive to moderate note segmentation errors, but its performance still depends on note-boundary accuracy. This behavior reflects a limitation of the sequential architecture, in which the downstream modules depend on the outputs of the note segmentator and the stationary pitch predictor.

\begin{table}[t]
\centering
\caption{Sensitivity of BERT-APC to note segmentation errors. We perturbed the predicted note boundaries by random displacements within $\pm 20$, $\pm 40$, $\pm 80$, and $\pm 160$ ms, and report the resulting RPA on the moderately and highly detuned subsets.}
\label{tab:boundary_perturbation}
\small
\setlength{\tabcolsep}{6pt}
\begin{tabular}{l|c|c}
\toprule
\textbf{Boundary Perturbation} & \textbf{\makecell{Moderately \\ Detuned (\%)}} & \textbf{\makecell{Highly \\ Detuned (\%)}} \\
\midrule
No perturbation & \textbf{94.95} & \textbf{89.24} \\
$\pm 20$ ms & 94.45 & 88.74 \\
$\pm 40$ ms & 93.55 & 87.94 \\
$\pm 80$ ms & 91.15 & 86.04 \\
$\pm 160$ ms & 85.85 & 81.74 \\
\bottomrule
\end{tabular}
\end{table}

\section{Conclusion}
In this study, we introduced a novel automatic pitch correction (APC) model, BERT-APC, that effectively corrects detuned singing voices without relying on any reference such as a music score, instrumental accompaniment, or guide vocals, while preserving intentional pitch deviations for expressive purposes. BERT-APC repurposes the symbolic music language model MusicBERT to predict note pitch sequences that are musically coherent from detuned pitch sequences. To this end, we developed a new deep learning-based note segmentator and a stationary pitch predictor, as well as a detuner for data augmentation.

In experiments, BERT-APC outperformed two recent singing voice transcription (SVT) models by large margins in note pitch prediction, and also received significantly higher MOS ratings compared to two widely-used commercial APC systems. 

One potential limitation of BERT-APC is that, since it corrects pitches based on musical context, its performance may degrade on songs that deviate significantly from typical musical patterns. One possible direction to mitigate this limitation is to incorporate additional musical context, such as symbolic accompaniment or song-scale information, when available. Such information could complement the vocal-derived context used by the current model and improve robustness on songs that are less well explained by typical musical patterns. Although using externally provided accompaniment information would move beyond the fully reference-free setting, a practical extension may be possible by automatically extracting such context from the audio signal itself (e.g., via Automatic Music Transcription).

To the best of our knowledge, BERT-APC is the first reference-free neural APC model that leverages a musical language model to correct detuned singing voices.

\bibliographystyle{IEEEtran}
\bibliography{references}

\newpage

\section{Biography Section}
 
\vspace{11pt}

\begin{IEEEbiography}[{\includegraphics[width=1in,height=1.25in,clip,keepaspectratio]{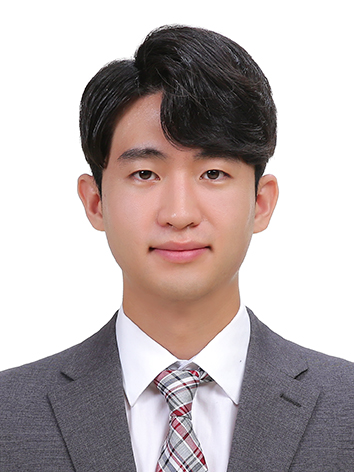}}]{Sungjae Kim}
received the B.S. and M.S. degrees in Computer Science and Electrical Engineering (CSEE) at Handong Global University. He is currently a Ph.D. student in CSEE at Handong Global University. Since 2019, he has been a student researcher at DL-LAB of Handong Global University, working under the supervision of Prof. Injung Kim. His research interests include deep learning, speech synthesis, singing voice synthesis, natural language processing, and speech recognition.
\end{IEEEbiography}

\begin{IEEEbiography}[{\includegraphics[width=1in,height=1.25in,clip,keepaspectratio]{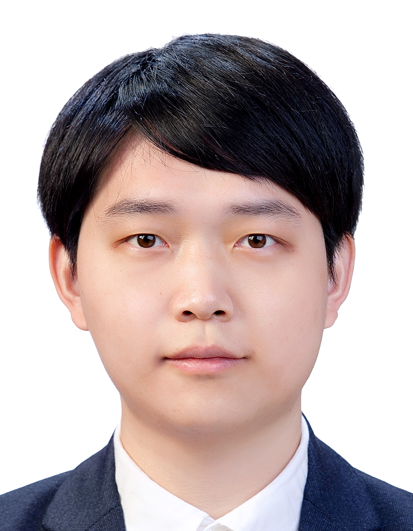}}]{Kihyun Na}
received a B.S. degree in Computer Science and Electrical Engineering (CSEE) from Handong Global University, Rep. of Korea, and an M.S. in Information and Communication Engineering from Ajou University, Rep. of Korea. He is currently pursuing a Ph.D. in CSEE at Handong Global University under the guidance of Prof. Injung Kim. His research interests include computer vision, generative models, and representation learning. 
\end{IEEEbiography}

\begin{IEEEbiography}[{\includegraphics[width=1in,height=1.25in,clip,keepaspectratio]{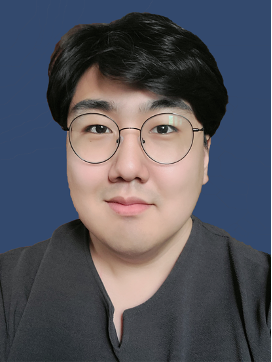}}]{Jinyoung Choi}
received B.S. and M.S. degrees in Computer Science and Electrical Engineering (CSEE) from Handong Global University, Rep. of Korea, and he is currently a Ph.D. student in CSEE at Handong Global University. Since 2021, he has been a student researcher at DL-LAB of Handong Global University, working under the supervision of Prof. Injung Kim. His research interests include deep learning, computer vision, and pattern recognition.
\end{IEEEbiography}

\begin{IEEEbiography}[{\includegraphics[width=1in,height=1.25in,clip,keepaspectratio]{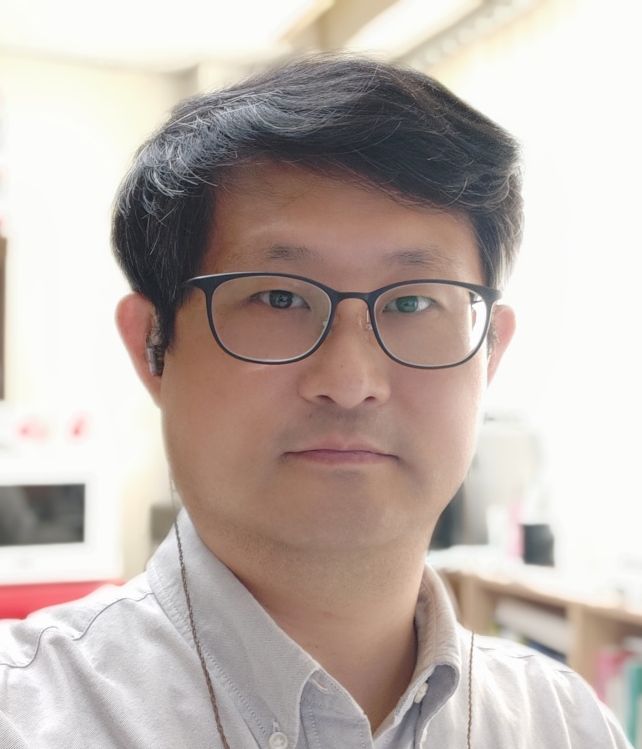}}]{Injung Kim}
is a professor of CSEE, Handong Global University since 2006.
He received B.S., M.S., and Ph.D. degrees in Computer Science from KAIST (Korea Advanced Institute of Science and Technology). He was a senior research engineer of Inzisoft. He was the Head of the School of CSEE, Program Director of the CS major, and an AI advisor of Samsung SW Center and POSCO, and currently, he is research advisor of multiple AI companies. His research interests include deep learning, image analysis and synthesis, speech synthesis, data analysis and prediction, recommendation system, outlier detection, and natural language processing.
\end{IEEEbiography}

\vspace{11pt}


\vfill

\end{document}